\documentclass[draftcls,onecolumn,11pt]{IEEEtran}
\usepackage{amsfonts,epsfig,amsmath,latexsym,amssymb,amscd,multirow,graphicx,geometry,lscape,amsmath,amssymb,bm,pifont,graphicx,amssymb,amsmath}
\usepackage{cancel,algorithm,algorithmic,algorithm,verbatim,array,multicol,palatino,amsfonts}
\usepackage{lineno}
\usepackage{subfig}
\usepackage[normalem]{ulem} 
\usepackage[dvips]{color}

\newcommand{\y}{\mathbf{y}}

\newcommand{\M}{\mathbf{M}}
\newcommand{\D}{\mathbf{D}}
\renewcommand{\r}{\mathbf{r}}

\newcommand{\w}{\mathbf{w}}

\newcommand{\I}{\mathbf{I}}

\newcommand{\X}{\mathbf{X}}
\newcommand{\s}{\mathbf{s}}
\newcommand{\f}{\mathbf{f}}

\newcommand{\Y}{\mathbf{Y}}
\newcommand{\R}{\mathbf{R}}
\newcommand{\F}{\mathbf{F}}
\newcommand{\W}{\mathbf{W}}

\newcommand{\V}{\mathbf{V}}
\newcommand{\G}{\mathbf{G}}
\renewcommand{\H}{\mathbf{H}}

\newcommand{\exE}{\mathbb{E}}

\newcommand{\GP}{\mathcal{GP}}
\newcommand{\THETA}{\bm{\theta}}

\newfont{\fsc}{eusm10}                         

\DeclareMathOperator*{\argmax}{arg\,max}
\newtheorem{theorem}{\noindent Theorem}

\modulolinenumbers[1]

\newcommand{\thickhline}{\noalign{\hrule height 0.8pt}}
\begin{document}
\title{System Identification in Wireless Relay Networks via Gaussian Process}
\author{Gareth W.~Peters$^{1,2}$, Ido Nevat$^{3}$, Jinhong Yuan$^{4}$ and Iain B. Collings$^{3}$ 
\begin{center}
{\footnotesize {\ 
\textit{$^{1}$ School of Mathematics and Statistics, University of New South Wales,
Sydney, 2052, Australia; \\[0pt]
$^{2}$ CSIRO  Sydney, Locked Bag 17, North Ryde, New South Wales, 1670, Australia \\[0pt]
$^{3}$ CSIRO, Wireless \& Networking Tech. Lab, Sydney, Australia.\\
$^{4}$ School of Electrical Engineering and Telecommunications, University of New South Wales, Sydney, Australia.\\
} } }
\end{center}
}
%
\maketitle

\begin{abstract}
\noindent We present a flexible stochastic model for a class of cooperative wireless relay networks, in which the relay processing functionality is not known at the destination. In addressing this problem we develop efficient algorithms 
to perform relay identification in a wireless relay network. We first construct a statistical model based on a representation of the system using Gaussian Processes in a non-standard manner due to the way we treat the imperfect channel state information. We then formulate the estimation problem to perform system identification, taking into account complexity and computational efficiency.
Next we develop a set of three algorithms to solve the identification problem each of decreasing complexity, trading-off the estimation bias for computational efficiency. The joint optimisation problem is tackled via a Bayesian framework using the Iterated Conditioning on the Modes methodology.
We develop a lower bound and several sub-optimal computationally efficient solutions to the identification problem, for comparison. We illustrate the estimation performance of our methodology for a range of widely used relay functionalities. The relative total error attained by our algorithm when compared to the lower bound is found to be at worst $9\%$ for low SNR values under all functions considered. The effect of the relay functional estimation error is also studied via BER simulations and is shown to be less than $2$ dB worse than the lower bound.  

\end{abstract}
\newpage
\section{Introduction to Relay System Identification}
Cooperative communications systems have been proposed to exploit the spatial diversity gains inherent in multiuser
wireless systems without the need of multiple antennas at each node, see \cite{nosratinia2004ccw} and \cite{laneman:2004}.
This is achieved by having the users relay each others messages and, thus, forming multiple transmission paths to the destination. Simply put, such a system broadcasts a signal from a transmitter at the source through a wireless channel. The signal
is then received by each relay node and a relay strategy is applied before the signal is retransmitted to the destination.
A number of relay strategies have been studied in the literature \cite{laneman:2004}, \cite{cover:1979}. The relay function can be optimized for different design objectives \cite{kramer:2005}, \cite{gomadam2007optimal}. For example, in the estimate and forward (EF) scheme, in the case of BPSK signaling, the optimal relay function is the hyperbolic tangent \cite{ghabeli2008new}. This relay function choice is optimal in this case as it has been shown to minimise the Mean Squared Error (MSE) at the relay \cite{gomadam2007optimal}. 
Other criteria for which the optimal relay function is non-linear include: capacity maximisation \cite{ghabeli2008new}, minimum error probability at the receiver \cite{crouse2007optimal}-\cite{yao2008sawtooth}, Signal-to-Noise (SNR) maximisation \cite{gomadam2007optimal} and rate maximisation \cite{cui2008some}. 

In general, the system identification problem may arise in several ways, see \cite{pillonetto2010new}, \cite{chiuso2011bayesian}, \cite{ko2009gp} and references within. For example, in an ad-hoc or sensor networks, it is possible that the destination does not have \textit{a priori} knowledge of the relay functionality utilised by each of the relays in the system \cite{volosencu2008identification}. Alternative scenarios that this problem may arise include networks which can adapt or cognitively learn suitable relay transmission functionality to optimise quality of service constraints such as capacity, throughput, bit-error-rate and transmission power \cite{khanpower}. 
These relay functions can be either static in time or in more advanced relay systems they may adapt over time, reacting and updating the transformations applied to the received signals before re-transmission to account for time varying channel characteristics.

Therefore, in all these scenarios, in order for the destination to perform detection of the data symbols, it first needs to perform estimation of the relay functionality.  This is a challenging problem due to the uncertain functional form of each relays processing on the received signal. To address this problem we introduce a class of semi-parametric modelling procedures based on Gaussian processes (GP), which are specifically designed to solve the identification problem.

GP define a family of stochastic processes that allow one to undertake flexible semi-parametric modelling of causal relationships without \textit{a priori} specification of the structure of the causal relationship. That is we may utilise a GP regression model as a flexible family of regression models, in which the relationship between the predictors (independent variables) and the responses (dependent variables) is not specified in advance. Instead, this relationship along with the parameters of this semi-parametric model are learnt jointly from the observed data. This is ideal in the context of system identification, where there is potential for highly non-linear relay function transformations. In this context, additional complications arise due to uncertain inputs to the relay function from the stochastic channel and thermal noise. We develop a stochastic model in which we define a distribution over a function space, corresponding to the class of continuous smooth functions $f(\cdot) \in \mathcal{C}$, which accounts for the uncertainty in the relay function. 

As discussed in tutorials by \cite{rasmussen2005gaussian}, \cite{mackay1998introduction} the Gaussian process literature is well established in spatial and temporal modelling, since the works of \cite{ripley1991statistical}. A few examples which include successful development of GP models in engineering include multi-user receiver design \cite{murillo2009gaussian}-\cite{perez2008digital}; channel equalisation and decoding \cite{olmos2010joint}; speech enhancement \cite{park2008gaussian}; source separation \cite{GPUSSTSP}; forecasting of non-linear dynamic systems \cite{girard2004approximate}; system identification of dynamical systems \cite{turner2009system}; and non-linear identification and equalization and separation of signals \cite{vaerenbergh2010kernel}.

\subsection{Contributions, Outline and Notation}
The main contributions of this work are as follows:
\begin{itemize}
	\item We propose a stochastic model for the problem of relay identification based on a flexible semi-parametric GP prior on the functional form of the relay processing functionality.
	\item We develop a formulation of this stochastic model under a sequential Bayesian estimation framework. In doing so we present three estimation scenarios and a comparative lower bound on their performance in order of complexity: 
	\begin{enumerate}
	\item{the first involves the optimal identification of the unknown relay function after integration of the nuisance parameters. We explain how this solution is computationally expensive for on-line solutions;}
	\item{the second involves an efficient, though sub-optimal algorithm. This is based on utilizing knowledge of pilot symbols in each transmission frame, zero forcing of the relay noise and imperfect channel state information;}
	\item{the third solution is based on the same assumptions as the second solution but under perfect channel state information;}
	\end{enumerate}
	\item Finally, we develop an estimation procedure based on Maximum A Posteriori (MAP) estimation jointly for the unknown relay function and unknown model parameters achieved via an adoption of the Iterated Conditioning on the Modes (ICM) methodology of \cite{besag1991bayesian} to the GP stochastic models developed.
\end{itemize}

\subsubsection{Outline}
The paper is organised as follows: in Section \ref{SystemDescription} we introduce a stochastic system model for the wireless relay system and the associated Bayesian model. In Section \ref{OptimalDetectionAlgorithm} we present the optimal identification algorithm. In Section \ref{System_Identification_Algorithms} we present a low complexity sub-optimal algorithms. Section \ref{simulationresults} presents results and analysis and conclusions are provided in Section \ref{conclusions}.
\subsubsection{Notation}
The following notation is used throughout: random variables are denoted by upper case letters and their realizations by lower case letters. In addition, bold will be used to denote a vector or matrix quantity, upper subscripts will refer to the relay node index and lower subscripts refer to the element of a vector or matrix.

\section{Background on Gaussian Processes}
In this section we present a brief outline on the statistical background for Gaussian Processes. A realisation of a random function from a GP can be considered simplistically as the analog of an infinite dimensional Gaussian random vector, in which each component of the vector corresponds to a point of evaluation of the unknown random function. Therefore, analogously to the multi-variate Gaussian random vector, the infinite dimensional GP prior is characterized sufficiently by a mean function and a covariance function. The covariance function must be specified carefully to ensure that finite realizations of the process correspond appropriately to multi-variate Gaussian random vectors. 

The main idea of utilizing a GP is to work with the unknown relay functions, without parameterizing this function. Instead, we place a GP prior directly on the space of functions that we wish to consider in the system identification.  The probabilistic nature of the GP model allows to define the space of admissible functions relating inputs to outputs, by simply specifying the mean and covariance functions of the process. 
For good introductions to GP see \cite{rasmussen2005gaussian}, \cite{mackay2003information}, \cite{seeger2004gaussian}. \\
\textbf{Definition 1 \cite{rasmussen2005gaussian}: }\textsl{ A Gaussian Process is a collection of random variables, any finite number of which have a joint Gaussian distribution.}

Furthermore, a GP is completely specified by the equivalent of sufficient statistics for a process, in this instance a mean function, denoted $m(\bm{x};\bm{\theta})$ and parameterised by $\bm{\theta}$, and a covariance function, denoted $\mathcal{C}\left(\bm{x},\bm{x}';\bm{\Psi}\right)$, parameterised by $\bm{\Psi}$ \cite{rasmussen2005gaussian}.

In the context of system identification, we formulate the problem as a semi-parametric regression model. In particular we encode our \textit{a-priori} belief in the functional form of the relay transform in terms of a prior distribution on a function space via a Gaussian process, denoted by the following prior 
\begin{equation}
f(\cdot) \sim \mathcal{GP}\left(m(\cdot;\bm{\theta}),\mathcal{C}\left(\cdot,\cdot;\bm{\Psi}\right)\right).
\end{equation}

Formally, this prior model ensures that for any finite set of predictor values or inputs to the unknown regression function $\left\{\bm{r}_t\right\}_{t=1:T}$, the corresponding random vector for the function at these points given by $\bm{f}_{1:T} = \left[f\left(\bm{r}_1\right),\ldots,f\left(\bm{r}_T\right)\right]$ is distributed according to the following multivariate Gaussian distribution,
\begin{equation}
\bm{f}_{1:T} \sim p\left(\bm{f}_{1:T}|\bm{R}_{1:T}=\bm{r}_{1:T}\right) = \mathcal{N}\left(\bm{f}_{1:T};  \left[m(\bm{r}_1;\bm{\theta}),\ldots,m(\bm{r}_T;\bm{\theta})\right],\mathcal{K}_{1:T}\right)
\end{equation}
with each component $\left[\mathcal{K}_{1:T}\right]_{ij,t} = \mathcal{C}\left(R_i(t),R_j(t)\right) = \mathbb{C}\text{ov}\left[f(R_i(t))f(R_j(t))\right]$.
Therefore we see that the covariance matrix, constructed from the kernel covariance function, measures the similarity between pairs of function values.
Therefore, as discussed in \cite{lazaro2010sparse} the elegance of a GP framework is that the properties of the unknown function to be estimated are expressed directly in terms of the covariance function, rather than implicitly via basis functions such as in a basis expansion model. To summarise this concept, we first consider two scalar inputs denoted by $R_i(t)$ and $R_j(t)$, separated by a distance of $||R_i(t)-R_j(t)||_2$, and we note that as we draw different realizations from the GP for function realizations at these points, $f(R_i(t))$ and $f(R_j(t))$, we will get different fluctuations depending on the function drawn. The degree to which this fluctuation in the values drawn occurs, is directly affected by the the choice of kernel. 

\section{Bayesian system model and Relay Identification} \label{SystemDescription}
\begin{figure}[h]
    \centering
        \epsfysize=8cm
        \epsfxsize=12cm
        \epsffile{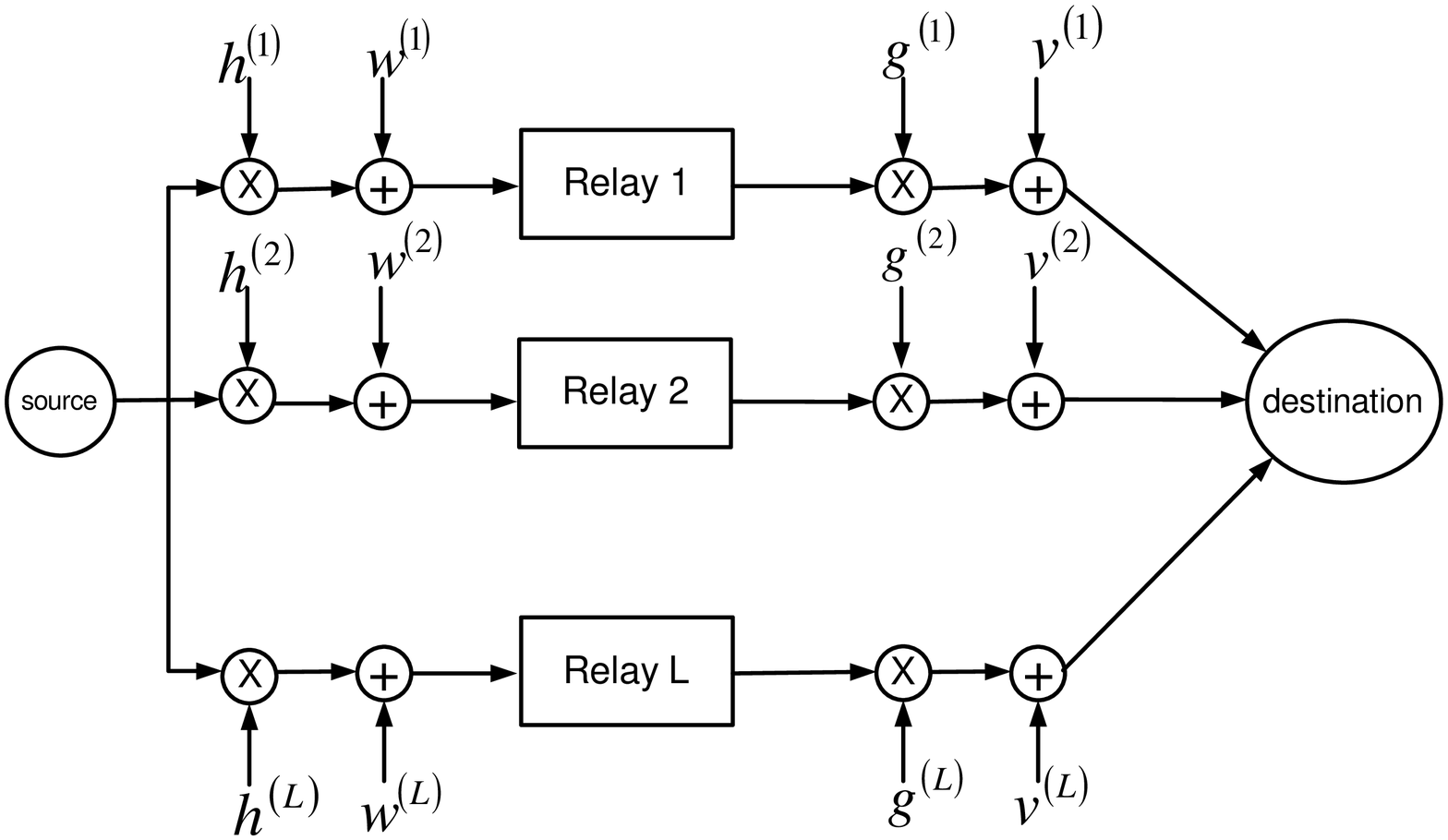}
        \caption{{\small{System model for transmission from one source, through $L$ relay channels to the destination.}}}
    \label{fig:system}
\end{figure}

We introduce the system model and a Bayesian model for inference on the system model parameters as depicted in Fig. \ref{fig:system}.
We will generically denote the frame index for the $t$-th frame using $t \in \left\{1,\ldots, T\right\}$. All the channels are modelled stochastically, where we do not know \textit{a priori} the realized channel coefficient values. Instead, we consider imperfect Channel State Information (CSI), in which we assume known statistics of the distribution of the channel coefficients.
\subsection{System model and assumptions} \label{System_model_and_assumptions}
\begin{figure}[b]
    \centering
        \epsfysize=8cm
        \epsfxsize=12cm
        \epsffile{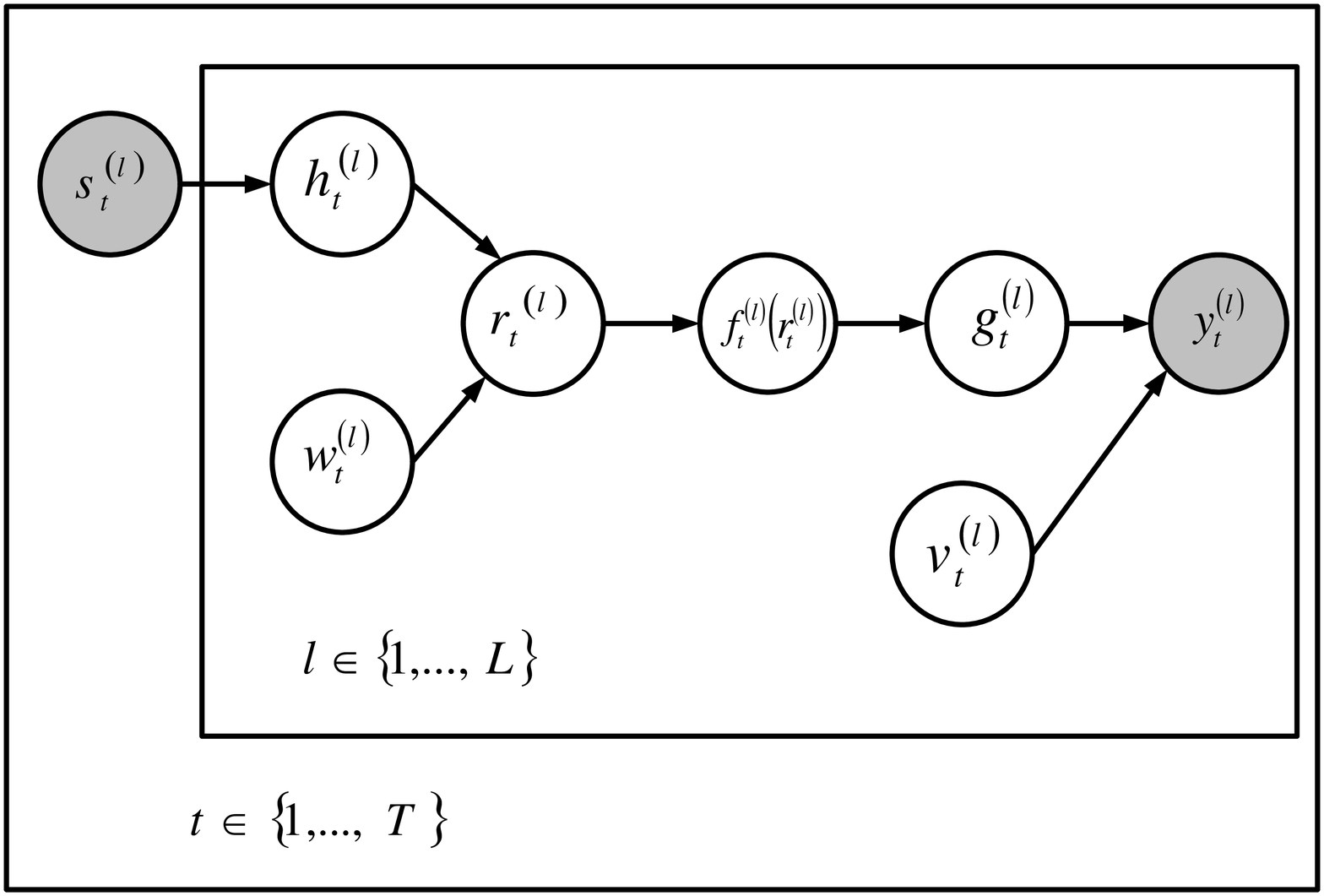}
        \caption{{\small{Directed Acyclic Graphical Model structure for the hierarchical Bayesian model.}}}
    \label{fig:DAG}
\end{figure}

\emph{
\begin{enumerate}
\item{Assume a wireless relay network with a single source node, transmitting sequences of $K$ pilot symbols per frame. We will denote this set of pilot symbols for frame $t$ as $\s=s_{1:K}$. These symbols are transmitted from a source to a single destination via $L$ relay nodes.}
\item{There are $L$ relays which cannot receive and transmit on the same time slot and on the same frequency band. We thus consider a half
duplex system model in which the data transmission is divided into
two steps. In the first step, the source node broadcasts a code
word $\s$ from the codebook to all the $L$ relay nodes. In
the second step, the relay nodes then transmit their signals
to the destination node on orthogonal non-interfering channels. We
assume that all channels are independent with a coherence interval
larger than the codeword length $K$.}
\item{Assume a general model for the CSI in which the estimates formed from the unknown realised channel coefficients for each relay link are known at the receiver {i.e. imperfect CSI}. 
This involves an assumption regarding the channel coefficients as follows:
\begin{itemize}
\item{From source to relay there are $L$ i.i.d. channels parameterized by
$\left\{H^{(l)} \sim
F\left(\widehat{h}^{(l)},\sigma_h^2\right)\right\}_{l=1}^L$, where
$F(\cdot)$ is the distribution of the channel coefficients, and the estimated channel, $\widehat{h}^{(l)}$.}
\item{From relay to destination there are $L$ i.i.d. channels parameterized by
$\left\{G^{(l)} \sim F\left(\widehat{g}^{(l)},\sigma_g^2\right)\right\}_{l=1}^L$, where
$F(\cdot)$ is the distribution of the channel coefficients, and the estimated channel, $\widehat{g}^{(l)}$.}
\end{itemize} }
\item{ The received signal at the $l$-th relay is a random vector given by
\begin{align}
\R^{(l)} = \s H^{(l)}+ \W^{(l)}, \; l \in
\left\{1,\ldots,L\right\},
\end{align}
where $H^{(l)}$ is the channel coefficient (scalar random variable) between the transmitter
and the $l$-th relay, $\s$ is the transmitted
code-word and $\W^{(l)}$ is the noise realization (vector random variable) associated with the relay.
\item{The transformation (relay function) of the received signal $\R^{(l)}$, performed by the $l$-th relay is assumed unknown. The unknown system model at each relay node $l$ will be modelled by a distribution over a function space as specified by a Gaussian Process prior, $$\f^{(l)}(\cdot) \sim \GP\left(\mu^{(l)}_{\THETA^{(l)}}\left(\cdot\right),\mathcal{C}^{(l)}_{\D^{(l)}}\left(\cdot,\cdot\right)\right).$$ 
Here $\f^{(l)}(\cdot)$ is defined to be the random vector function (relay function). A realization of this random vector function will be denoted by $\f^{(l)}(\R^{(l)})=\left[f^{(l)}(R_1^{(l)}),\ldots,  f^{(l)}(R_K^{(l)})   \right]$, which is evaluated for the received signal at the $l$-th relay. The distribution of possible functions to be considered is controlled by the GP mean function $\mu^{(l)}_{\THETA^{(l)}}(\cdot)$ parameterised by $\THETA^{(l)}$ and covariance function $k^{(l)}_{\D^{(l)}}\left(\cdot,\cdot\right)$ constructed from a kernel parametrised by $\D^{(l)}.$\\
We denote time series observations of the function evaluation at the $l$-th relay
\begin{align}
\begin{split}
&\f^{(l)}_{1:T} = \left(
\underbrace{f^{(l)}\left(R^{(l)}_1\left(1\right)\right),\ldots, f^{(l)}\left(R^{(l)}_K\left(1\right)\right)}_{t=1},
\underbrace{f^{(l)}\left(R^{(l)}_1\left(2\right)\right),\ldots, f^{(l)}\left(R^{(l)}_K\left(2\right)\right)}_{t=2},
\right.
\\&\hspace{1cm}
\left.\ldots,
\underbrace{f^{(l)}\left(R^{(l)}_1\left(T\right)\right),\ldots, f^{(l)}\left(R^{(l)}_K\left(T\right)\right)}_{t=T}
\right)^T.
\end{split}
\end{align}
}}
\item{To ensure a parsimonious and estimatable statistical model, particularly when L is large, we assume that all relay functions will have the same class of mean and covariance functions.}
\item{ 
We consider the following model structure for the relay functionality:
\begin{itemize}
	\item The choice of mean function considered will be restricted to linear constants and trend models of the form $\mu^{(l)}_{\bm{\theta}^{(l)}}(R_k^{(l)}) = \theta_1^{(l)} + \theta_2^{(l)} R_k^{(l)}$. This assumption is consistent with the forms of relay function considered in the literature such as the Amplify-and-Forward (AF) of \cite{zhao2006improving} or the Estimate-and-Forward (EF) of \cite{gomadam2007optimal}.
	\item The kernel function, $\mathcal{K}^{(l)}$ of the $l$-th relay given in (\ref{KernelMatrix}), can be expressed as the following squared exponential model:
\begin{align}	 
\label{kernel_function}
\mathcal{C}(R_i^{(l)},R_j^{(l)}) = \exp\left(-\frac{||R_i^{(l)}-R_j^{(l)}||^2}{2 d^2}\right),
\end{align}
$\forall i,j \in \left\{1,\ldots, K\right\}$.
This widely used kernel produces smooth functions with the properties that the covariance function is stationary and non-degenerate \cite{rasmussen2005gaussian}. Using this kernel, the corresponding covariance matrix for the $l$-th relay can be expressed as
\begin{align}
\mathcal{K}^{(l)}_{1:T}=
\left[
\begin{array}{ccc}
 \mathcal{C}\left(R_1^{(l)}(1),R_1^{(l)}(1)\right) &  \cdots    & \mathcal{C}\left(R_1^{(l)}(1),R_K^{(l)}(T)\right)\\
 \vdots &  \ddots    & \vdots\\
 \mathcal{C}\left(R_K^{(l)}(T),R_1^{(l)}(1)\right) &  \cdots    & \mathcal{C}\left(R_K^{(l)}(T),R_K^{(l)}(T)\right)
\end{array}
\right],
\label{KernelMatrix}
\end{align}
\item The corresponding mean vector for the $l$-th relay can be expressed as
\begin{align}
\begin{split}
\bm{\mu}^{(l)}_{1:T} = 
&\left(
\underbrace{
\mu^{(l)}_{\theta^{(l)}}\left(R_1^{(l)}(1)\right),\ldots,
\mu^{(l)}_{\theta^{(l)}}\left(R_K^{(l)}(1)\right)
}_{t=1},
\underbrace{\mu^{(l)}_{\theta^{(l)}}\left(R_1^{(l)}(2)\right),\ldots, \mu^{(l)}_{\theta^{(l)}}\left(R_K^{(l)}(2)\right)
}_{t=2},
\ldots,\right.\\
&\left.
\underbrace{\mu^{(l)}_{\theta^{(l)}}\left(R_1^{(l)}(T)\right),\ldots, \mu^{(l)}_{\theta^{(l)}}\left(R_K^{(l)}(T)\right)
}_{t=T}
\right)^T
\end{split}
\end{align}
\item Conditional on the mean functions and covariance functions, $\mu^{(l)}_{\THETA^{(l)}}\left(\cdot\right),\mathcal{C}^{(l)}_{\D^{(l)}}\left(\cdot,\cdot\right)$ and $\mu^{(m)}_{\THETA^{(m)}}\left(\cdot\right),\mathcal{C}^{(m)}_{\D^{(m)}}\left(\cdot,\cdot\right)$	we consider realisations of each Gaussian Process function to be statistically independent. Therefore we are assuming the following model structure:
	$$\mathbb{E}\left[\f^{(l)}\left(\X\right) \f^{(m)}\left(\Y\right)^{T} |\mu^{(l)}_{\THETA^{(l)}}\left(\cdot\right),\mathcal{C}^{(l)}_{\D^{(l)}}\left(\cdot,\cdot\right), \mu^{(m)}_{\THETA^{(m)}}\left(\cdot\right),\mathcal{C}^{(m)}_{\D^{(m)}}\left(\cdot,\cdot\right)\right]=\bm{\Sigma}$$
for all $\X,\Y$ inputs and all relays $l,m$, where $\bm{\Sigma}$ is a diagonal covariance matrix. This gives us spatial independence between the functionality of each relay.
\end{itemize}
}
\item{Conditional on matrix $\f = \left(\f^{(1)}(R^{(1)}),\ldots,\f^{(L)}(R^{(L)})\right)$, the received signal at the destination, from the $l$-th relay, is a random vector given by
\begin{align}
\label{system_model}
\Y^{(l)} = \f^{(l)}\left(\R^{(l)}\right)G^{(l)}+\V^{(l)}, \; l\in
\left\{1,\ldots,L\right\},
\end{align}
where the scalar random variable $G^{(l)}$ is the channel coefficient between the $l$-th
relay and the receiver, \\ $\f^{(l)}\left(\r^{(l)}\right)
\triangleq
\left[f^{(l)}\left(r_1^{(l)}\right),\ldots,f^{(l)}\left(r_K^{(l)}\right)
\right]^{T}$ is the memoryless relay processing function (with
possibly different functions at each of the relays) and the random vector $\V^{(l)}$
is the noise realization associated with the receiver.\\
We define
\begin{align}
\begin{split}
&\Y_{1:T}= \left(\Y^{(1)}_{1:T},\ldots, \Y^{(L)}_{1:T}\right), \text{where}\\
&\Y^{(l)}_{1:T} = \left(
\underbrace{Y^{(l)}_1\left(1\right),\ldots, Y^{(l)}_K\left(1\right)}_{t=1},
\underbrace{Y^{(l)}_1\left(2\right),\ldots, Y^{(l)}_K\left(2\right)}_{t=2},
\ldots,
\underbrace{Y^{(l)}_1\left(T\right),\ldots, Y^{(l)}_K\left(T\right)}_{t=T}
\right)^T.
\end{split}
\end{align}
}
\item{All received signals are corrupted by zero-mean additive white complex
Gaussian noise. At the $l$-th relay the noise corresponding to the
$l$-th transmitted symbol is denoted by random variable
$W_{i}^{(l)} \sim \mathcal{CN}\left(0,\sigma_w^2\right)$. At the
receiver this is denoted by random variable $V_{i}^{(l)}
\sim \mathcal{ CN}\left(0,\sigma_v^2\right)$. Additionally, we assume the following properties:
\begin{equation*}
\mathbb{E}\left[W_{i}^{(l)} \overline{W}_{j}^{(m)} \right]=
\mathbb{E}\left[V_{i}^{(l)} \overline{V}_{j}^{(m)} \right]=
\mathbb{E}\left[W_{i}^{(l)} \overline{V}_{j}^{(m)} \right]=0,
\end{equation*}
$\forall i,j \in \left\{1,\ldots,K\right\}, \forall l,m \in
\left\{1,\ldots,L\right\}, i \neq j, l \neq m$, where
$\overline{W}_{j}$ denotes the complex conjugate of $W_j$.\\
We define
\begin{align}
\begin{split}
&\W_{1:T}= \left(\W^{(1)}_{1:T},\ldots, \W^{(L)}_{1:T}\right), \text{where}\\
&\W^{(l)}_{1:T} = \left(
\underbrace{W^{(l)}_1\left(1\right),\ldots, W^{(l)}_K\left(1\right)}_{t=1},
\underbrace{W^{(l)}_1\left(2\right),\ldots, W^{(l)}_K\left(2\right)}_{t=2},
\ldots,
\underbrace{W^{(l)}_1\left(T\right),\ldots, W^{(l)}_K\left(T\right)}_{t=T}
\right)^T.
\end{split}
\end{align}
}
\end{enumerate}
}
We now present the posterior parameters required to be estimated, followed by the prior distributional choices, finishing with the posterior distribution for the directed acyclic graph in Fig. \ref{fig:DAG}. 
Given the posterior distribution we formulate the relay system identification problem in Section \ref{OptimalDetectionAlgorithm}. 
The posterior parameters and functions of interest are given by the parameter vector after observing $T$ frames,
\begin{align}
\begin{split}
&\left(\f\left(\cdot\right), \bm{\theta}, \D\right)= \left(\f^{(1)}\left(\cdot\right),\ldots, \f^{(L)\left(\cdot\right)}, \bm{\theta}^{(1:L)}, \D^{(1:L)}\right).
\end{split}
\end{align}
The prior choices for the relay functionality for $\f$ are given by a GP with hyper priors for the mean and covariance functions as specified below. \\
\textbf{Prior Model Structure}
\emph{
\begin{enumerate}
\item{The priors of the hyper-parameters associated with the linear mean function are given by 
$\bm{\theta} = \left[\bm{\theta^{(1)}},\ldots,\bm{\theta^{(L)}} \right]$, with 
$\bm{\theta^{(l)}} = \left[\theta^{(l)}_1, \theta^{(l)}_2 \right]$, where 
$\theta^{(l)}_1 \sim N(0,1)$, $\theta^{(l)}_2 \sim N(0,100)$ for all $l$. Note, here we assume a vague prior for the gradient of the mean function.}
\item{The priors of the hyper-parameters, associated with the kernel function $\mathcal{C}$ in (\ref{kernel_function}), considered in the construction of the covariance function is specified by
 $\D = \left[D^{(1)},\ldots, D^{(L)}\right]$, where $D^{(i)}\sim U\left[0, 10\right]$.
}
\item{The hierarchical prior for the $l$-th relay function is then given by $\bm{f}^{(l)}(\cdot) \sim \GP\left(\mu^{(l)}_{\THETA^{(l)}}\left(\cdot\right),\mathcal{C}^{(l)}_{\D^{(l)}}\left(\cdot,\cdot\right)\right) $, with GP mean function $\mu^{(l)}_{\THETA^{(l)}}$ parameterised by $\THETA^{(l)}$ and covariance function $\mathcal{C}^{(l)}_{\D^{(l)}}\left(\cdot,\cdot\right)$ constructed from a kernel parametrised by $\D^{(l)}$.}
\end{enumerate}
}
\noindent \textbf{Posterior Model Structure}\\
The combination of the likelihood model, priors for model parameters and symbols, and the hierarchical priors for the GP prior, when combined under Bayes' Theorem, results in a full posterior distribution: 
\begin{equation}
\begin{split}
&p\left( \THETA,\D,\w^{1:L}_{1:T}, \f^{1:L}\left(\cdot\right)|\y_{1:T}^{1:L} \right)\\
&\propto
\prod_{l=1}^L p\left(\y_{1:T}^{(l)}|\f^{(l)}_{1:T},\THETA^{(l)},D^{(l)},\w^{(l)}_{1:T}\right)
\GP\left(\f^{(l)}\left(\cdot\right);\mu^{(l)}_{\THETA^{(l)}}\left(\cdot\right),\mathcal{C}^{(l)}_{D^{(l)}}\left(\cdot,\cdot\right)\right) p\left(\w^{(l)}_{1:T}\right) p\left(\THETA^{(l)}\right) p\left(D^{(l)}\right)\\
&=
\prod_{l=1}^L
\left(
\prod_{t=1}^T
p\left(\y^{(l)}(t)|\f^{(l)}\left(\r^{(l)}(t)\right),\THETA^{(l)},D^{(l)},\w^{(l)}(t)\right)
\GP\left(\f^{(l)}\left(\cdot\right);\mu^{(l)}_{\THETA^{(l)}}\left(\cdot\right),\mathcal{C}^{(l)}_{D^{(l)}}\left(\cdot,\cdot\right)\right) \right.\\
& \left.\times p\left(\w^{(l)}(t)\right) \right)p\left(\THETA^{(l)}\right) p\left(D^{(l)}\right)\\
&=
\prod_{l=1}^L
\left(
\prod_{t=1}^T
\prod_{k=1}^K
p\left(y_k^{(l)}(t)|f^{(l)}\left(r_k^{(l)}(t)\right),\THETA^{(l)},D^{(l)},w_k^{(l)}(t)\right)
\GP\left(\f^{(l)}\left(\cdot\right);\mu^{(l)}_{\THETA^{(l)}}\left(\cdot\right),\mathcal{C}^{(l)}_{D^{(l)}}\left(\cdot,\cdot\right)\right) \right.\\
& \left.\times p\left(\w_k^{(l)}(t)\right) \right)p\left(\THETA^{(l)}\right) p\left(D^{(l)}\right).
\end{split}
\label{PosteriorModel}
\end{equation}
Note, we have included auxiliary parameters $\w^{1:L}_{1:k}$ to represent the unknown noise realizations at the $L$ relays for each transmitted sequence of symbols. The augmentation of these auxiliary parameters in the posterior specification allows us to obtain closed form expressions for the likelihood model, in particular a Gaussian form which will be relevant when combined with the GP prior for the relay functionality. Without the introduction of these auxiliary nuisance parameters we would be unable to derive a closed form expression for the relay function likelihoods, see discussions in \cite{peters2010bayesian}. 

\section{Problem Formulation for System Identification and Inference} \label{OptimalDetectionAlgorithm}
Now we can specify the marginal posterior distributions of particular interest to the identification problem using the Bayesian model presented in (\ref{PosteriorModel}). We note that typically in the standard GP regression framework, the values of the covariate, in our model given by $r_k^{(l)}(t)$, are known or observed inputs, one corresponding to each of the observations of the symbols at each of the relays. However, in our model, these are random and unobserved variables due to the additive noise at the relay. In the following section we specify three different solutions to the estimation problem.

\subsection{Problem Formulation I}
\noindent
The most general framework is based on the posterior estimation of the GP mean and covariance functions, parameterized by the parameters $\bm{\theta},\D$. \\
\textbf{System Identification Definition 1:}
\begin{equation}
\begin{split}
&\left\{\widehat{\f}^{(1:L)}\left(\cdot\right),
\widehat{\THETA}^{(1:L)},\widehat{D}^{(1:L)}
\right\} = \\
&\argmax_{\f^{(1:L)}\left(\cdot\right), \THETA^{(1:L)}, D^{(1:L)}}
\int 
\prod_{l=1}^L
\left(
\prod_{t=1}^T
\prod_{k=1}^K
d w_k^{(l)}(t)
p\left(y_k^{(l)}(t)|f^{(l)}\left(r_k^{(l)}(t)\right),\THETA^{(l)},D^{(l)},w_k^{(l)}(t)\right)
p\left(\w_k^{(l)}(t)\right) 
\right)\\
&\times \GP\left(\f^{(l)}\left(\cdot\right);\mu^{(l)}_{\THETA^{(l)}}\left(\cdot\right),\mathcal{C}^{(l)}_{D^{(l)}}\left(\cdot,\cdot\right)
 \right)p\left(\THETA^{(l)}\right) p\left(D^{(l)}\right).
 \end{split}
\label{Identification_problem1}
\end{equation}
This solution is optimal from the perspective that it minimises the variance of the estimator due to the Rao-Blackwellisation performed to integrate the nuisance parameters corresponding to the relay noise. This problem has a high computational complexity and would typically be solved via a numerical procedure, such as the Markov Chain Monte Carlo (MCMC). Unfortunately, MCMC solutions would not easily adapt in this scenario to a computationally efficient solution, see \cite{peters2010bayesian}.

We focus on an alternative class of solutions that has a much lower computational complexity, which admit recursive estimation algorithms. Instead of performing numerical integration followed by numerical optimization to perform relay identification, we solve the following problem via an iterative optimisation procedure.
\subsection{Problem Formulation II}
\noindent
Formulating computationally efficient estimation procedure, we shall make the following assumptions.\\
\textbf{Assumption I:} \textit{Consider imperfect CSI. We condition inference on a noisy estimate of the sufficient statistics of the channels, given by $\exE[\G^{(l)}]=\widehat{g}^{(l)}$, $\exE[\H^{(l)}]=\widehat{h}^{(l)}$.}\\
\textbf{Assumption II:} \textit{We consider zero forcing (ZF) condition for the relay thermal noise given by $\W^{(l)}=\exE[\W^{(l)}]=0$.}\\
As a result of these assumptions, the received signal at the relay is given by
\begin{align}
\label{ZF_model}
\begin{split}
\r_{\text{ZF}}^{(l)}|\left(\s, \H^{(l)}=\widehat{h}^{(l)}, \W^{(l)}=0\right) &= \s \widehat{h}^{(l)}, \; l \in \left\{1,\cdots,L\right\},\\
\Y^{(l)}|\left(\s, \H^{(l)}=\widehat{h}^{(l)} ,\G^{(l)}=\widehat{g}^{(l)}, \W^{(l)}=0 \right)  &= \f^{(l)}\left(\s \widehat{h}^{(l)}\right)\widehat{g}^{(l)}+ \V^{(l)}, \; l\in \left\{1,\cdots,L\right\},
\end{split}
\end{align}
where  $\f^{(l)}\left(\cdot\right)\sim GP\left(\mu^{(l)}_{\THETA^{(l)}}\left(\cdot\right),\mathcal{C}^{(l)}_{\D^{(l)}}\left(\cdot,\cdot\right)\right)$.\\
\textbf{System Identification Definition 2:} \textsl{Under assumptions I and II, the resulting identification problem can be stated as the following
\small
\begin{equation}
\begin{split}
&\left\{\widehat{\f}^{(1:L)}\left(\cdot\right)
\widehat{\THETA}^{(1:L)},\widehat{D}^{(1:L)}
\right\} = \\
&\argmax_{\f^{(1:L)}\left(\cdot\right), \THETA^{(1:L)}, D^{(1:L)}}
\prod_{l=1}^L
\left(
\prod_{t=1}^T
\prod_{k=1}^K
p\left(y_k^{(l)}(t)|f^{(l)}\left(r_k^{(l)}(t)\right)
 \right)
 \right)
\GP\left(\f^{(l)}\left(\cdot\right);\mu^{(l)}_{\THETA^{(l)}}\left(\cdot\right),\mathcal{C}^{(l)}_{D^{(l)}}\left(\cdot,\cdot\right)\right)
  p\left(\THETA^{(l)}\right) p\left(D^{(l)}\right).
 \end{split}
\label{Identification_problem2}
\end{equation}}
\subsection{Problem Formulation III}
\noindent
As a quantification of the best case identification accuracy we present the following lower bound estimation, based on the following assumptions. \\
\textbf{Assumption III:} \textit{Consider perfect CSI where all the channels, denoted by $\G^{(l)}$ and $H^{(l)}$, are known exactly at the receiver.}\\
Through Assumption II and Assumption III, we obtain a lower bound on the accuracy and performance that one can achieve. This is because they result in knowledge of the following form,
\begin{align}
\label{lower_bound}
\begin{split}
\r_{\text{ZF}}^{(l)}|\left(\s, \H^{(l)}=h^{(l)}, \W^{(l)}=0\right) &= \s h^{(l)}, \; l \in \left\{1,\ldots,L\right\},\\
\Y^{(l)}|\left(\s, \H^{(l)}=h^{(l)} ,\G^{(l)}=g^{(l)}, \W^{(l)}=0 \right)  &= \f^{(l)}\left(\s h^{(l)}\right) g^{(l)}+ \V^{(l)}, \; l\in \left\{1,\ldots,L\right\}.
\end{split}
\end{align}
This allows us to make the following third system identification definition for the lower bound on performance.\\
\textbf{System Identification Definition 3:} \textsl{Under assumptions II and III, the resulting lower bound on the identification problem can be stated as the following
\begin{equation}
\begin{split}
&\left\{\widehat{\f}^{(1:L)}_{LB}\left(\cdot\right),
\widehat{\THETA}^{(1:L)}_{LB},
\widehat{D}_{LB}^{(1:L)}
\right\} = \\
&\argmax_{\f^{(1:L)}\left(\cdot\right), \THETA^{(1:L)}, D^{(1:L)}}
\prod_{l=1}^L
\left(
\prod_{t=1}^T
\prod_{k=1}^K
p\left(y_k^{(l)}(t)|f^{(l)}\left(r_k^{(l)}(t)\right) \right)\right) 
\GP\left(\f^{(l)}\left(\cdot\right);\mu^{(l)}_{\THETA^{(l)}}\left(\cdot\right),\mathcal{C}^{(l)}_{D^{(l)}}\left(\cdot,\cdot\right)\right)
  p\left(\THETA^{(l)}\right) p\left(D^{(l)}\right).
 \end{split}
\label{Identification_problem3}
\end{equation}}

We note that the estimation problem involves estimation of the Maximum a Posteriori (MAP) relay function values locally at each of these predictor locations given by the MAP estimator $\widehat{\f}^{(l)}(\cdot)$. One may then utilise the results of this estimation for prediction of the unknown mean structure of the function $f^{(l)}(\cdot)$, conditional on estimated hyper parameters for the mean and covariance function, at new input values of the received signal, $r_*$, via the identity
\begin{equation}
\label{GP_estimation}
 \overline{f_*}^{(l)} \triangleq \exE
 \left[
f^{(l)}(r_*)|\y^{(l)}_T,\widehat{\bm{\theta}}^{(l)},\widehat{\D}^{(l)},\widehat{\R}^{(l)}_T,r_*\right] = \mathcal{K}^{(l)}(r_*,\widehat{\R}^{(l)}_T) \left(\mathcal{K}^{(l)}_T + \sigma_v^2\I \right)^{-1}\y^{(l)}_T.
 \end{equation}
and the associated estimation error variance
\begin{equation}
\begin{split}
\label{GP_estimation_error}
 \sigma_*^{(l)} \triangleq &\exE
 \left[
 \left(f^{(l)}(r_*)-\overline{f_*}^{(l)}\right)^2 
 |\y^{(l)}_T,\widehat{\bm{\theta}}^{(l)},\widehat{\D}^{(l)},\widehat{\R}^{(l)}_T,r_*
 \right] \\
 =& 
 \mathcal{K}^{(l)}(r_*,r_*)-
 \mathcal{K}^{(l)}(r_*,\widehat{\R}^{(l)}_T) \left(\mathcal{K}^{(l)}_T + \sigma_v^2\I \right)^{-1}
 \left(\mathcal{K}^{(l)}(r_*,\widehat{\R}^{(l)}_T)\right)^T.
\end{split}
\end{equation}
\section{System Identification via Iterated Conditional Modes} \label{System_Identification_Algorithms}
Having formulated the relay identification problem, we now address algorithmic procedures to solve efficiently (\ref{Identification_problem1}). 
\subsection{Understanding Iterated Conditioning on the Modes Estimation for System Identification}
ICM was originally developed in \cite{lindley1972bayes} and \cite{besag1991bayesian} for efficient MAP estimation in very high dimensional Bayesian models such as Markov random fields which were used in the analysis of images with speckle noise. Since these papers, ICM procedures have been developed in many other areas of estimation, see examples in \cite{chen2008unsupervised} and \cite{zhang2008markov}. 

As discussed in \cite{rowe2003multivariate} ICM is fundamentally a deterministic optimization method that finds the joint posterior modal estimators corresponding to the MAP estimates. We illustrate this simple algorithm on a generic two parameter example developed in \cite{rowe2003multivariate} before extending the concept to the solution to the relay identification problem. In this example a generic bivariate posterior distribution $p(\theta_1,\theta_2|y_{1:T})$ is considered and the aim is to find the MAP estimate corresponding to the mode, which satisfies
\begin{equation}
\frac{\partial}{\partial \theta_1} p(\theta_1,\theta_2|y_{1:T})|_{\theta_2 = \widehat{\theta}_2}=\frac{\partial}{\partial \theta_2} p(\theta_1,\theta_2|y_{1:T})|_{\theta_1 = \widehat{\theta}_1}=0.
\end{equation}
This is equivalent to 
\begin{equation}
p(\theta_2|y_{1:T})\frac{\partial}{\partial \theta_1} p(\theta_1|\theta_2,y_{1:T})|_{\theta_2 = \widehat{\theta}_2}=p(\theta_1|y_{1:T})\frac{\partial}{\partial \theta_2} p(\theta_2|\theta_1,y_{1:T})|_{\theta_1 = \widehat{\theta}_1}=0,
\end{equation}
assuming that $p(\theta_1|y_{1:T}) \neq 0$ and $p(\theta_2|y_{1:T}) \neq 0$. Hence, given the full conditional posterior distributions $p(\theta_1|\theta_2,y_{1:T})$ and $p(\theta_2|\theta_1,y_{1:T})$ and solutions for their modes $\widehat{\theta}_1 = \widehat{\theta}_1(\theta_2,y_{1:T})$ and $\widehat{\theta}_2 = \widehat{\theta}_2(\theta_1,y_{1:T})$, the algorithm then proceeds by initializing the mode estimates and successively iterating over updates of conditional mode estimates. For example at iteration $j$, the update of $\widehat{\theta}^{(j)}_1$ conditions on the mode estimate of $\widehat{\theta}^{(j-1)}_2$ and then the update of $\widehat{\theta}^{(j)}_2$ conditions on the mode estimate $\widehat{\theta}^{(j)}_1$. This is repeated either for a fixed number of iterations or until a convergence criterion is satisfied. When the posterior full conditionals are unimodal, this procedure is guaranteed to converge to the global maximum. In other multi-modal settings, the ICM algorithm is guaranteed to converge to a maximum, though it may be a local maximum \cite{besag1991bayesian}.

Hence, all variants of the ICM estimation procedure have the following in common: they first involve specification of a multivariate posterior distribution, deconstructed as a set of full conditional posterior distributions. These full conditional posterior distributions are given, for the $l$-th relay, based on the full posterior in (\ref{PosteriorModel})
{\small{
\begin{subequations}
\begin{align}
\label{posterior_1}
&p\left( \f^{l}\left(\cdot\right)|\THETA,\D,\y_{1:T}^{(1:L)} \right) \propto \prod_{t=1}^T \prod_{k=1}^K
p\left(y_k^{(l)}(t)|f^{(l)}\left(r_k^{(l)}(t)\right),\THETA^{(l)},D^{(l)},w_k^{(l)}(t)\right)
\GP\left(\f^{(l)}\left(\cdot\right);\mu^{(l)}_{\THETA^{(l)}}\left(\cdot\right),\mathcal{C}^{(l)}_{D^{(l)}}\left(\cdot,\cdot\right)\right);\\
\label{posterior_2}
&p\left( \THETA|\D, \f^{l}\left(\cdot\right),\y_{1:T}^{(l)} \right) \propto \prod_{t=1}^T \prod_{k=1}^K \GP\left(\f^{(l)}\left(\cdot\right);\mu^{(l)}_{\THETA^{(l)}}\left(\cdot\right),\mathcal{C}^{(l)}_{D^{(l)}}\left(\cdot,\cdot\right)\right) p\left(\THETA^{(l)}\right);\\
\label{posterior_3}
&p\left( \D|\THETA, \f^{l}\left(\cdot\right),\y_{1:T}^{(l)} \right) \propto \prod_{t=1}^T \prod_{k=1}^K \GP\left(\f^{(l)}\left(\cdot\right);\mu^{(l)}_{\THETA^{(l)}}\left(\cdot\right),\mathcal{C}^{(l)}_{D^{(l)}}\left(\cdot,\cdot\right)\right) p\left(D^{(l)}\right).
\end{align}
\end{subequations}
}}
Next, given these full conditional posterior distributions, and observations for the $l$-th relay, applying the ICM algorithm involves initializing the estimate of the MAP solutions, denoted by $\left(\widehat{\f}^{l,(0)}_{1:T},\widehat{\THETA}^{(0)},\widehat{\D}^{(0)}\right)$. Then the $j$-th iteration of the ICM algorithm successively updates each estimate of $\left(\widehat{\f}^{l,(j)}_{1:T},\widehat{\THETA}^{(j)},\widehat{\D}^{(j)}\right)$
based on the solutions at iteration $(j-1)$, and the solutions to the following sequence of MAP estimates for the full conditional posteriors,
\begin{subequations}
\begin{align}
\label{posterior_solution_1}
&\text{Vec}[\widehat{\f}^{l,(j)}_{1:T}] = \argmax_{\f^{l}_{1:T}} p\left( \f^{l}\left(\cdot\right)|\widehat{\THETA}^{(j-1)},\widehat{\D}^{(j-1)},\y_{1:T}^{(l)} \right);\\
\label{posterior_solution_2}
&\widehat{\THETA}^{(j)} = \argmax_{\THETA} p\left( \THETA|\widehat{\D}^{(j-1)}, \widehat{\f}^{l,(j)}_{1:T},\y_{1:T}^{(l)} \right);\\
\label{posterior_solution_3}
&\widehat{\D}^{(j)} = \argmax_{\D} p\left( \D|\widehat{\THETA}^{(j)}, \widehat{\f}^{l,(j)}_{1:T},\y_{1:T}^{(l)} \right).
\end{align}
\end{subequations}
Repeating this procedure can guarantee convergence to a maximum, see \cite{besag1991bayesian}, who noted that ICM can be made equivalent to instantaneous freezing in a stochastic optimization procedure known as simulated annealing \cite{kirkpatrick1983optimization}.
Iterating this procedure successively produces a sequence of MAP estimates which converge to an optimum solution, typically for a small number of iterations $J$. Additionally, the number of ICM iterations $J$ required to obtain the optimal MAP solutions for each set of $T$ frames of $K$ symbols in our problem formulation is designed to be independent of $T$ and $K$. This is due to the conjugacy we exploit for the model MAP estimation of the vector component which grows linearly in dimension with $KT$, corresponding to the vector $\text{Vec}[\f^{l}_{1:T}]$. 
\subsection{Generic ICM Estimation for Wireless Relay System Identification via GP's}
To develop an ICM algorithm, we need to construct a block Gibbs framework for posterior inference, and obtain expressions for the mode of the full conditional posterior distributions. This is equivalent to finding the conditional MAP estimate. In this problem we will exploit conjugacy properties of the posterior model in (\ref{Identification_problem1}). 
\begin{theorem}
\label{Theorem_1}
\textsl{The full conditional distributions under (\ref{ZF_model}) are given by:
\begin{enumerate}
	\item The full conditional posterior distribution for the $l$-th relay function in (\ref{posterior_1}) is given by
\begin{equation}
p\left( \text{Vec}\left[\f^{(l)}_{1:T} \right]| \THETA^{(l)},D^{(l)},\y^{(l)}_{1:T} \right)
= N\left(\M, \Sigma \right),
\end{equation}
where 
\begin{equation}
\M = \left(
\left(\mathcal{K}^{(l)}_{1:T}\right)^{-1} + \frac{1}{\sigma_{\V}^2}\I_{K \times T}\right)^{-1}
		\left(\left(\mathcal{K}^{(l)}_{1:T}\right)^{-1} \M_0 
		+ \frac{1}{\sigma_{\V}^2} \text{Vec}\left[\y^{(l)}_{1:T} \right]   \right),
\label{MeanFunction}		
\end{equation}
with 
\begin{equation}
\M_0 = \left(
\underbrace{\mu^{(l)}_{\THETA^{(l)}}\left(r^{(l)}_1(1)\right),\ldots,\mu^{(l)}_{\THETA^{(l)}}\left(r^{(l)}_K(1)\right)}_{t=1}, \ldots,
\underbrace{\mu^{(l)}_{\THETA^{(l)}}\left(r^{(l)}_1(T)\right),\ldots,\mu^{(l)}_{\THETA^{(l)}}\left(r^{(l)}_K(T)\right)}_{t=T}
\right)^T
\end{equation}
and 
\begin{equation}
\label{Conjugate_2}
 \Sigma =  \left(
\left(\mathcal{K}^{(l)}_{1:T}\right)^{-1} + \frac{1}{\sigma_{\V}^2}\I_{K \times T}\right)^{-1}.
\end{equation}
The conditional MAP estimate of the relay function in (\ref{posterior_solution_1}), denoted by $\text{Vec}\left[\widehat{\f}^{(l)}_{1:T} \right]$, is given by
\begin{equation}
 \text{Vec}\left[\widehat{\f}^{(l)}_{1:T} \right] = \argmax_{\f^{(l)}_{1:T}}
 p\left( \text{Vec}\left[\f^{(l)}_{1:T} \right]| \THETA^{(l)},D^{(l)},\y^{(l)}_{1:T} \right)
 = \M.
\end{equation}
\item 
The full conditional for $\THETA^{(l)}$ in (\ref{posterior_2}) can be expressed as
\begin{equation}
p\left( \THETA^{(l)}|\D^{(l)}, \f^{(l)}_{1:T}\left(\cdot\right),\y^{(l)}_{1:T} \right)
\propto
N \left(\text{Vec}\left[\f^{(l)}_{1:T}\right];\bm{\mu}^{(l)}_{1:T},\mathcal{K}^{(l)}_{1:T}\right)
p\left(\THETA^{(l)}\right).
\end{equation}
The conditional MAP estimate of $\THETA^{(l)} = \left[\theta^{(l)}_1, \theta^{(l)}_2 \right]$ in (\ref{posterior_solution_2}), denoted by $\widehat{\THETA}^{(l)}$, is given by
\begin{equation}
\begin{split}
\theta_1 &= \frac
{-\theta_2\left(
\left[\left(\Sigma_{\bm{\Theta}}\right)^{-1}\right]_{2 ,1}+
\sum_{i=1}^K\sum_{t=1}^Tr_i(t) \left[\left(\mathcal{K}_{1:T}\right)^{-1}\right]_{i ,t}  \right)
-\frac{1}{2} \sum_{i=1}^K \sum_{t=1}^T f(r_i(t)) \left[\left(\mathcal{K}_{1:T}\right)^{-1}\right]_{i ,t}}
{
\left[\left(\Sigma_{\bm{\Theta}}\right)^{-1}\right]_{1 ,1}+
\sum_{i=1}^K\sum_{t=1}^T r_i(t) \left[\left(\mathcal{K}_{1:T}\right)^{-1}\right]_{i ,t}  
}\\
\theta_2 &= \frac{1}{2 \Phi}
\left(
-\sum_{i=1}^K\sum_{t=1}^T r_i(t)  f(r_i(t)) \left[\left(\mathcal{K}_{1:T}\right)^{-1}\right]_{i ,t}
\right.\\
&\left.+
\frac
{
\left(
\sum_{i=1}^K\sum_{t=1}^T f(r_i(t)) \left[\left(\mathcal{K}_{1:T}\right)^{-1}\right]_{i ,t}
\right)
\left(
\left[\left(\Sigma_{\bm{\Theta}}\right)^{-1}\right]_{1 ,2}
+
\sum_{i=1}^K\sum_{t=1}^T r^2_i(t) \left[\left(\mathcal{K}_{1:T}\right)^{-1}\right]_{i ,t}
\right)
}
{
\left[\left(\Sigma_{\bm{\Theta}}\right)^{-1}\right]_{1 ,1}+
\sum_{i=1}^K\sum_{t=1}^T r_i(t) \left[\left(\mathcal{K}_{1:T}\right)^{-1}\right]_{i ,t}  
}
\right),
\end{split}
\end{equation}
where 
\begin{equation}
\begin{split}
\Phi = 
&\frac{-\left(
\left[\left(\Sigma_{\bm{\Theta}}\right)^{-1}\right]_{2 ,1}+
\sum_{i=1}^K\sum_{t=1}^T r_i(t) \left[\left(\mathcal{K}_{1:T}\right)^{-1}\right]_{i ,t}  
\right)
\left(
\left[\left(\Sigma_{\bm{\Theta}}\right)^{-1}\right]_{1 ,2}+
\sum_{i=1}^K\sum_{t=1}^T r^2_i(t) \left[\left(\mathcal{K}_{1:T}\right)^{-1}\right]_{i ,t}  
\right)
}
{
\left[\left(\Sigma_{\bm{\Theta}}\right)^{-1}\right]_{1 ,1}+
\sum_{i=1}^K\sum_{t=1}^T r_i(t) \left[\left(\mathcal{K}_{1:T}\right)^{-1}\right]_{i ,t}  
}\\
&+
\left[\left(\Sigma_{\bm{\Theta}}\right)^{-1}\right]_{2 ,2}+
\sum_{i=1}^K\sum_{t=1}^T r^2_i(t) \left[\left(\mathcal{K}_{1:T}\right)^{-1}\right]_{i ,t}  .
\end{split}
\end{equation}
\item 
The full conditional for $\D^{(l)}$ in (\ref{posterior_3}) can be expressed as
\begin{equation}
\begin{split}
p\left(\D^{(l)} | \f^{(l)}_{1:T},\THETA^{(l)},\y^{(l)}_{1:T} \right)
\propto
N \left(\text{Vec}\left[\f^{(l)}_{1:T}\right];\bm{\mu}^{(l)}_{1:T},\mathcal{K}^{(l)}_{1:T}\right)
p\left(\D^{(l)}\right).
\end{split}
\end{equation}
The conditional MAP estimate of $D^{(l)}$ in (\ref{posterior_solution_3}), denoted by $\widehat{D}^{(l)}$, is given as the numerical solution to
\small
\begin{equation}
\left[
-\frac{1}{2} \text{Tr}\left(
\mathcal{K}_{1:T}^{-1}\frac{d\mathcal{K}_{1:T}}{d D^{(l)}}
\right)
+ \frac{1}{2}
\left(\text{Vec}\left[\f_{1:T}\right]^T-\bm{\mu}_{1:T}\right)^T
\left(\mathcal{K}_{1:T}\right)^{-1}
\frac{d\mathcal{K}_{1:T}}{d D^{(l)}}
\left(\mathcal{K}_{1:T}\right)^{-1}
\left(\text{Vec}\left[\f_{1:T}\right]^T-\bm{\mu}_{1:T}\right)
\right]
\mathbb{I}\left[D^{(l)} \in \left(a,b\right)\right]=0,
\end{equation}
\normalsize
where $\mathbb{I}\left[\cdot\right]$ is the indicator function, and we have 
\begin{equation}
\frac{d\mathcal{K}_{1:T}}{d D^{(l)}} = \frac{1}{D^{(l)} }\Phi_{1:T} \odot \mathcal{K}_{1:T},
\end{equation}
and $\odot$ is the Kronecker product, and the $i,j$-th element of $\Phi_{1:T}$ is given by
$\left[\Phi_{1:T}\right]_{i,j} \triangleq  ||R_i^{(l)}-R_j^{(l)}||^2$.
\end{enumerate}
}
\end{theorem}
The proof of Theorem \ref{Theorem_1} is provided in Appendix 1.
\subsection{Recursive ICM-GP System Identification Algorithm}
We develop a version of Iterative Conditioning on the Mode (ICM) \cite{besag1993statistical} which exploits different structural properties of the GP model. 

We begin with the most computational approach, though it represents the optimal solution as it utilises all information obtained to update the estimated function.\\
\noindent \textbf{Online Estimation I - Full Information}\\
This approach is optimal since, for each relay $l$, having observed $t$-frames of $K$-symbols of observed data $Y^{(l)}_{1:t}=\left(Y_1^{(l)}(1),\ldots,Y_K^{(l)}(1),\ldots,Y_1^{(l)}(t),\ldots,Y_K^{(l)}(t)\right)$, we evaluate $\mathcal{K}^{(l)}_{1:t}$ based on 
\begin{align}
\mathcal{K}^{(l)}_{1:t}=
\left[
\begin{array}{ccc}
 \mathcal{C}\left(R_1^{(l)}(1),R_1^{(l)}(1)\right) &  \cdots    & \mathcal{C}\left(R_1^{(l)}(1),R_K^{(l)}(t)\right)\\
 \vdots &  \ddots    & \vdots\\
 \mathcal{C}\left(R_K^{(l)}(t),R_1^{(l)}(1)\right) &  \cdots    & \mathcal{C}\left(R_K^{(l)}(t),R_K^{(l)}(t)\right)
\end{array}
\right].
\end{align}
This approach utilises all the observed information in the estimation of the ICM algorithm. However for $J$ iterations of the ICM algorithm it will be of complexity $O(JK^2t^2)$ in memory usage and $O(JK^3t^3)$ in computational complexity. This is primarily due to the cost of inverting the Gram matrix in each update stage of the ICM algorithm. However, in this approach each estimation is performed based on a new frame of length $K$ of observed symbols and the estimation of the matrix of function values, $\bm{f}^{(l)}_{1:t}$, is completely updated based on all past history. For example, the evaluation of (\ref{MeanFunction}) for updating the ICM MAP estimate of the function values is given by,
\begin{equation}
\M = \left(
\left(\mathcal{K}^{(l)}_{1:T}\right)^{-1} + \frac{1}{\sigma_{\V}^2}\I_{K \times T}\right)^{-1}
		\left(\left(\mathcal{K}^{(l)}_{1:t}\right)^{-1} \M_0 
		+ \frac{1}{\sigma_{\V}^2} \text{Vec}\left[\y^{(l)}_{1:t} \right]   \right).
\end{equation}
This expression involves the evaluation of $\mathcal{K}^{(l)}_{1:t}$ and its inverse, which become rapidly computationally expensive though analytically exact for the update of the MAP estimate of the function values $\bm{f}^{(l)}_{1:t}$ at each iteration of the ICM algorithm. We circumvent this growing computational complexity with two alternative approaches presented next and compare their performance to this optimal solution.\\
\noindent \textbf{Online Estimation II - Frame-by-Frame}\\
This approach is sub-optimal though significantly more computationally efficient. In this proposed approach, for each relay $l$, having observed the $t$-th frame of $K$-symbols $Y^{(l)}_{t}=\left(Y_1^{(l)}(t),\ldots,Y_K^{(l)}(t)\right)$, we evaluate $\mathcal{K}^{(l)}_{1:t}$ based on just the current frame of data,
\begin{align}
\mathcal{K}^{(l)}_{t}=
\left[
\begin{array}{ccc}
 \mathcal{C}\left(R_1^{(l)}(t),R_1^{(l)}(t)\right) &  \cdots    & \mathcal{C}\left(R_1^{(l)}(t),R_K^{(l)}(t)\right)\\
 \vdots &  \ddots    & \vdots\\
 \mathcal{C}\left(R_K^{(l)}(t),R_1^{(l)}(t)\right) &  \cdots    & \mathcal{C}\left(R_K^{(l)}(t),R_K^{(l)}(t)\right)
\end{array}
\right].
\end{align}
Effectively it is equivalent to partitioning $\mathcal{K}^{(l)}_{1:t}$ as follows,
\begin{equation*}
\begin{split}
\mathcal{K}^{(l)}_{1:t} &= \mathcal{K}^{(l)}_{1} \oplus \mathcal{K}^{(l)}_{2} \oplus \cdots \oplus \mathcal{K}^{(l)}_{t} 
=\left[
\begin{array}{cccc}
\mathcal{K}^{(l)}_{1} 	& \bm{0} 		& \ldots 	& \bm{0} \\
\bm{0} 		& \mathcal{K}^{(l)}_{2} 	& \ldots 	& \bm{0} \\
\vdots 		& \ddots 		& \ddots 	& \vdots \\
\bm{0} 		& \bm{0}  	&	\bm{0} & \mathcal{K}^{(l)}_{t} \end{array}
\right].
\end{split}
\end{equation*}
However, for $J$ iterations of the ICM algorithm at frame $t$ it will be of complexity $O(JK^2)$ in memory usage and $O(JK^3)$ in computational complexity. In addition, since the relay functions being estimated are not time varying, the estimates obtained on previous frames ($\f^{(l)}_{1:t-1}$) and their uncertainty can be combined in several different approaches for example via an update according to the following recursions, 
\begin{equation}
\begin{split}
\bm{m}_{t} &= \bm{m}_{t-1} + \frac{1}{t}\left(\tilde{\f}^{(l)}_{t} - \bm{m}_{t-1}\right)\\
\Phi_{t} &= \Phi_{t-1} + \frac{1}{t}\left(\left(\tilde{\f}^{(l)}_{t} - \bm{m}_{t}\right)\left(\tilde{\f}^{(l)}_{t} - \bm{m}_{t}\right)' - \Phi_{t-1}\right).
\end{split}
\label{RecursiveMeanCov}
\end{equation}
Here, the matrix of function estimates, denoted by $\tilde{\f}^{(l)}_{t}$, corresponds to function values obtained for frame $t$ from the ICM estimates $\f^{(l)}_{t}$, quantised to a grid of predictor values $R_{1},\ldots,R_{S}$ which partition the convex hull of the relay symbols based on the constellation being transmitted. These quantised values are then included in an average of the function values for each frame, where the mean function vector at these quantised grid points is denoted by $\bm{m}_{t}$ and its uncertainty measured by a covariance in the estimate at these quantised values is denoted by $\Phi_{t}$.\\
\noindent \textbf{Online Estimation III - Sliding Window}\\
This approach provides a trade-off in computational complexity and optimality between the proposed estimation approaches I and II. It is sub-optimal, though significantly more computationally efficient than approach I and it recursively uses previous frames as opposed to the block wise analysis in approach II. 

Under this proposed approach, for each relay $l$, having observed $\tau$ symbols of observed data  $Y^{(l)}_{\tau}=\left(Y_1^{(l)}(1),\ldots,Y_K^{(l)}(1),\ldots,Y_k^{(l)}(\tau)\right)$, we evaluate $\mathcal{K}^{(l)}_{1:t}$ according to a sliding window based on the past block of $S$ observed symbols. For example we may consider blocks of length $K$ (i.e. length of the frame) and an update of the function estimates for each newly observed symbols observation. This corresponds to utilizing a Gram-matrix for each update with structure given by,
\begin{align}
\mathcal{K}^{(l)}_{\tau} &= 
\left[
\begin{array}{cc}
 \bar{\mathcal{K}}^{(l)}_{\tau-1} &  \bar{\bm{k}}^{(l)}_{\tau-1}\left(R^{(l)}(\tau)\right)\\
 \bar{\bm{k}}^{(l)}_{\tau-1}\left(R^{(l)}(\tau)\right)^T & \mathcal{C}\left(R^{(l)}(\tau),R^{(l)}(\tau)\right)
\end{array}
\right] ,
\end{align}
where we have dropped the frame index on the received signals which is redundant in this specification. We define the last row and column of the new matrix as the vector\\ $\bar{\bm{k}}^{(l)}_{\tau-1}\left(R^{(l)}(\tau)\right) = \left[\mathcal{C}\left(R^{(l)}(\tau-S+1),R^{(l)}(\tau)\right),\ldots,\mathcal{C}\left(R^{(l)}(\tau-1),R^{(l)}(\tau)\right)\right]$. In addition the modified matrix $\bar{\mathcal{K}}^{(l)}_{\tau-1}$ corresponds to the ``down sized'' version of the regularised matrix $\mathcal{K}^{(l)}_{\tau-1}$ obtained by removing the first row and column, ensuring the sliding window structure and the fixed dimensionality of $\mathcal{K}^{(l)}_{\tau}$ for all times $\tau $.

This particular approach is computationally efficient as it allows one to utilise a special computational evaluation of the inverse matrix $\mathcal{K}^{(l)}_{\tau}$ in (\ref{MeanFunction}) when updating the ICM MAP estimates. This involves utilising recursive knowledge of the previous updates matrix and its inverse $\left[\mathcal{K}^{(l)}_{\tau-1}\right]^{-1}$ under an approach described in \cite{vaerenbergh2010kernel}. This approach utilises the Sherman - Morrison - Woodbury matrix inversion Lemma \cite{petersen2008matrix} under two applications detailed below.
\begin{enumerate}
\item{Evaluate the inverse of the down sized matrix $\bar{\mathcal{K}}^{(l)}_{\tau-1}$.}
\item{Evaluate the inverse of the up sized matrix $\mathcal{K}^{(l)}_{\tau}$ based on the inverse of $\bar{\mathcal{K}}^{(l)}_{\tau-1}$.}
\end{enumerate}

Dropping the relay index $l$ for convenience, the inverse of the down sized matrix is obtained using the previous evaluation of the matrix, decomposed according to
\begin{align}
\mathcal{K}_{\tau-1} &= 
\left[
\begin{array}{cc}
 \mathcal{C}\left(R(\tau-S+1),R(\tau-S+1)\right)   & \bar{\bm{k}}_{\tau-1}\left(R(\tau-1)\right)\\
\bar{\bm{k}}_{\tau-1}\left(R(\tau-1)\right)  & \bar{\mathcal{K}}_{\tau-1}
\end{array}
\right], 
\end{align}
and its inverse decomposed as
\begin{align}
\mathcal{K}^{-1}_{\tau-1} &= 
\left[
\begin{array}{cc}
 \left[\mathcal{K}^{-1}_{\tau-1}\right]_{11}   &  \left[\mathcal{K}^{-1}_{\tau-1}\right]_{12:1S}\\
\left[\mathcal{K}^{-1}_{\tau-1}\right]^{T}_{12:1S}  & \left[\mathcal{K}^{-1}_{\tau-1}\right]_{22:SS}
\end{array}
\right],
\end{align}
where $\left[\mathcal{K}^{-1}_{\tau-1}\right]_{22:SS}$ 
denotes the lower $(2 \times S)(2 \times S)$ sub-block of the inverse matrix $\mathcal{K}^{-1}_{\tau-1}$. Under these decompositions, the inverse for the down sized matrix is obtained using the following matrix inversion lemma identity \cite{sherman1949adjustment} 
$$\bar{\mathcal{K}}^{-1}_{\tau-1} =  \left[\mathcal{K}^{-1}_{\tau-1}\right]_{22:SS} - \frac{1}{\left[\mathcal{K}^{-1}_{\tau-1}\right]_{11}}\left[\mathcal{K}^{-1}_{\tau-1}\right]_{12:1S}\left[\mathcal{K}^{-1}_{\tau-1}\right]^T_{12:1S}.$$

Having obtained the inverse of the down sized matrix in terms of values known at iteration $\left(\tau - 1\right)$, the inverse of the upsized matrix proceeds according to the following decompositions again via application of the matrix inversion lemma. Again, dropping the relay index $l$ for convenience, the inverse of the up sized matrix is obtained using the previous evaluation of the matrix, decomposed according to
\begin{align}
\mathcal{K}_{\tau} &= 
\left[
\begin{array}{cc}
\bar{\mathcal{K}}_{\tau-1}  & \bar{\bm{k}}_{\tau}\left(R(\tau)\right)\\
\bar{\bm{k}}_{\tau}\left(R(\tau)\right)^T  &  \mathcal{C}\left(R(\tau),R(\tau)\right)
\end{array}
\right], 
\end{align}
and its inverse decomposed as
\begin{align}
\mathcal{K}^{-1}_{\tau} &= 
\left[
\begin{array}{cc}
 \left[\mathcal{K}^{-1}_{\tau}\right]_{11:(S-1)(S-1)}   &  \left[\mathcal{K}^{-1}_{\tau}\right]_{12:1S}\\
\left[\mathcal{K}^{-1}_{\tau}\right]^{T}_{12:1S}  & \left[\mathcal{K}^{-1}_{\tau}\right]_{SS}
\end{array}
\right],
\end{align}
where $\left[\mathcal{K}^{-1}_{\tau}\right]_{11:(S-1)(S-1)}$ denotes the upper $(2 \times S)(2 \times S)$ sub-block of the inverse matrix $\mathcal{K}^{-1}_{\tau}$. Applying the matrix inversion lemma, one can obtain further decomposition according to the previously evaluated down sized matrix inverse as follows,
\begin{align}
\mathcal{K}^{-1}_{\tau} &= 
\left[
\begin{array}{cc}
 \bar{\mathcal{K}}^{-1}_{\tau-1}\left(I + \bar{\bm{k}}_{\tau}\left(R(\tau)\right)\bar{\bm{k}}_{\tau}\left(R(\tau)\right)^T\bar{\mathcal{K}}^{-1H}_{\tau-1}\left[\mathcal{K}^{-1}_{\tau}\right]_{SS}\right)    &  -\bar{\mathcal{K}}^{-1}_{\tau-1}\bar{\bm{k}}_{\tau}\left(R(\tau)\right)\left[\mathcal{K}^{-1}_{\tau}\right]_{SS}\\
-\left(\bar{\mathcal{K}}^{-1}_{\tau-1}\bar{\bm{k}}_{\tau}\left(R(\tau)\right)\right)^T\left[\mathcal{K}^{-1}_{\tau}\right]_{SS}  & \left[\mathcal{K}^{-1}_{\tau}\right]_{SS}
\end{array}
\right],
\end{align}
with $\left[\mathcal{K}^{-1}_{\tau}\right]_{SS} = \left(\mathcal{C}\left(R(\tau),R(\tau)\right) - \bar{\bm{k}}_{\tau}\left(R(\tau)\right)^T\bar{\mathcal{K}}^{-1}_{\tau-1}\bar{\bm{k}}_{\tau}\left(R(\tau)\right) \right)^{-1}$.

In addition, since the relay functions being estimated are not time varying, as detailed in Approach II, the estimates obtained on previous frames ($\f^{(l)}_{1:\tau-1}$) and their uncertainty can be combined in several different approaches for example via the update mechanism described in (\ref{RecursiveMeanCov}).

\section{Simulation Results}  
\label{simulationresults}
\begin{figure}
  \centering
  \subfloat[Frame-by-Frame]
  {
  \includegraphics[width=7.5cm, height=6cm]{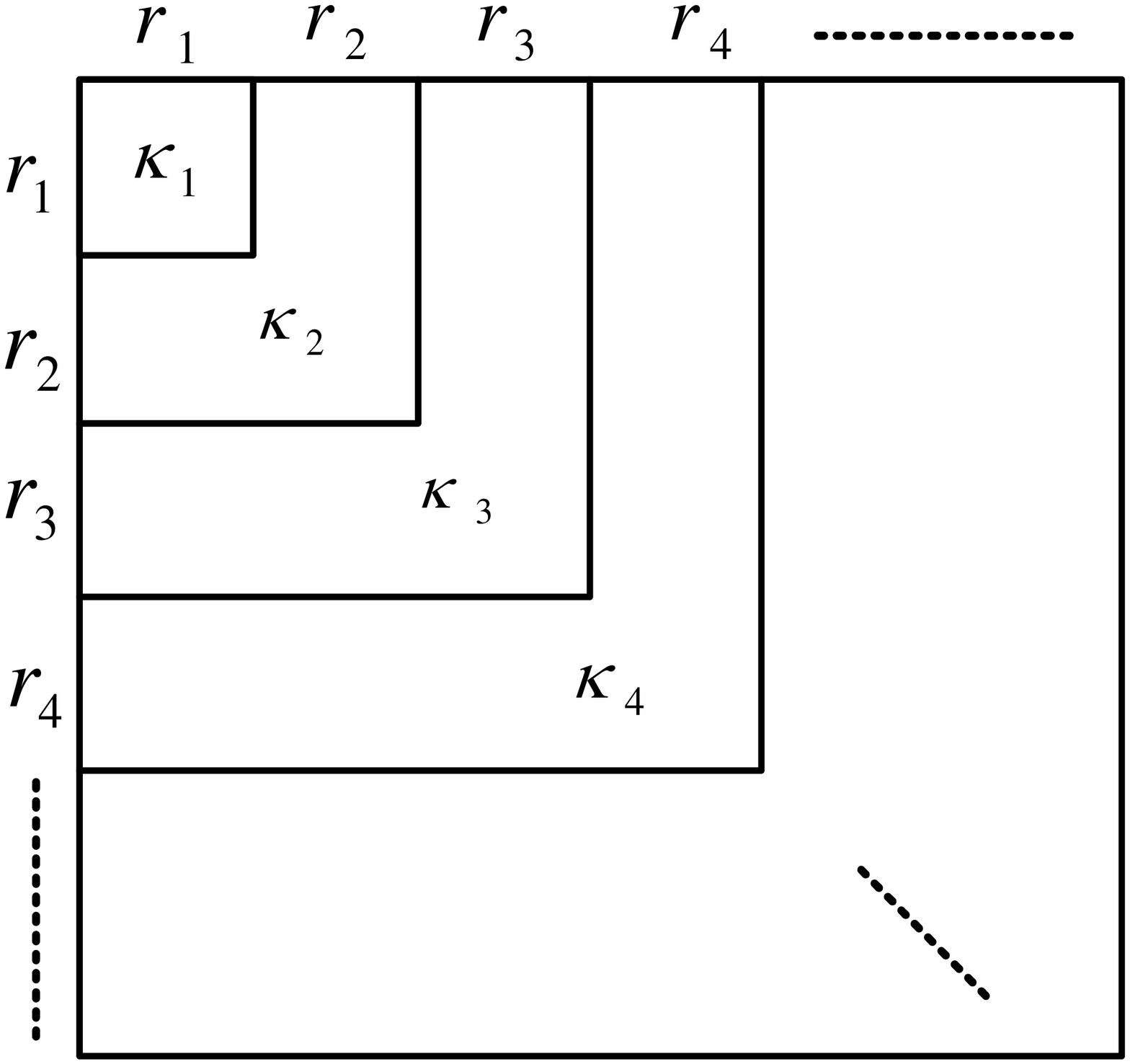}
  }  
  \subfloat[Sliding window]
  {\includegraphics[width=7.5cm, height=6cm]
  {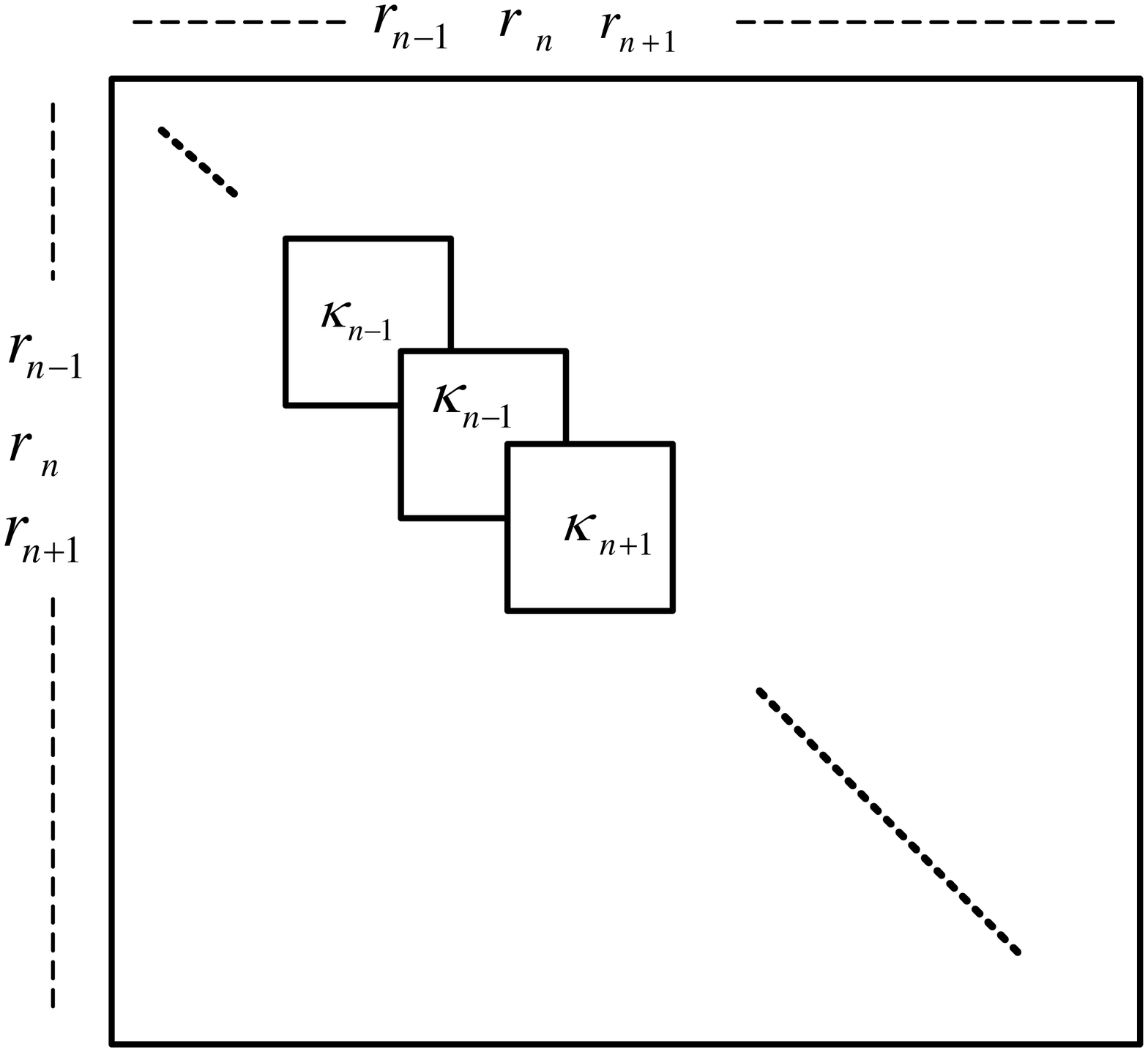}}
\caption{\small{Construction of Kernel matrix under Frame-by-Frame (Approach II) and Sliding Window (Approach III). }}
  \label{fig:animals}
\end{figure}

In this section, we present the performance of the proposed algorithms via Monte Carlo simulations.
The simulation settings for all the simulations are as follows:
\begin{itemize}
	\item The prior distribution for all the channels is Rayleigh fading, and the channels are assumed to be both spatially and temporally independent;
\item The channels uncertainty was set to $\sigma^2_{\G}=\sigma^2_{\F}=0.2$;
\item We define the \textit{received SNR} as the ratio of the average received signal power to the average noise power,
\begin{align*}
\text{SNR} \triangleq 
10 \log \frac
{\text{Tr}\left[\exE\left[\left(\G(l) \F(l) s(l)\right)\left(\G(l) \F(l) s(l)\right)^H\right]\right]}
{\text{Tr}\left[\exE\left[\left(\G(l) \V(l) + \W(l)\right)\left(\G(l) \V(l) +\W(l)\right)^H\right]\right]}=
10 \log \frac
{1}
{\sigma^2_{\V}+\sigma^2_{\W}};
\end{align*}
\item The SNR is set to $0$ dB (low SNR) and $10$ dB (high SNR);
\item The results are obtained from simulations over $T=100$ transmitted frames with $K=200$ symbols per frame; 
\item In all simulations 16PAM constellations were considered;
\item The ICM algorithm iterated $J=50$ times over the solutions to (\ref{posterior_solution_1}-\ref{posterior_solution_3});
\item The relay functions tested correspond to: \textit{absolute value} $f(x) = |x|$, \textit{linear (affine transformation)} $f(x) = a x + b$, \textit{hyperbolic-tan} $f(x) = a \tanh(w x + \phi)$ and \textit{demodulated};
\item Simulations were performed for full kernel matrix estimation Approach I under both perfect CSI and imperfect CSI; partial kernel matrix estimation Approach II with frame by frame estimation under both perfect CSI and imperfect CSI; and a sliding window with $50\%$ overlap estimation Approach III under both perfect CSI and imperfect CSI.
\end{itemize}
We begin with an analysis for the case in which the algorithm is applied to an increasing number of frames of transmitted symbols, according to the Full Information (Approach I) in which the kernel matrix is not truncated in any manner. These are the results with the highest computational complexity but without any reduction of the kernel matrix, hence these results were produced form the full set of observed information over time, as presented diagrammatically in Subplot (a) in Fig. \ref{fig:animals}. 

\begin{figure}
    \centering
        \epsfysize=7cm
        \epsfxsize=17cm
        \epsffile{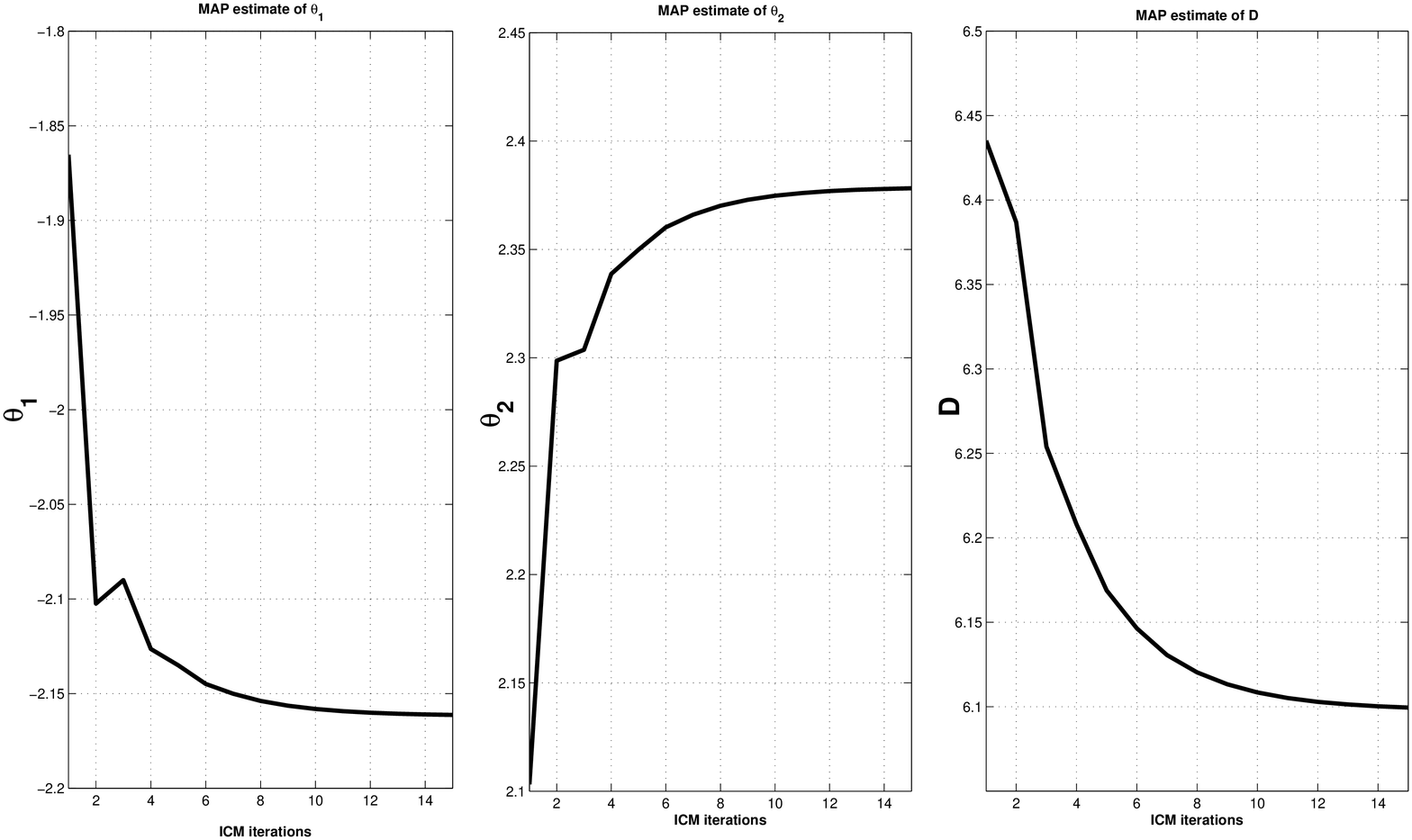}
    \caption{Convergence of hyperparameters estimates $\theta^{(l)}_1$, $\theta^{(l)}_2$ and $\D$}
  \label{fig:hyperparameters}
\end{figure}

Analysis of the performance of the ICM algorithm can be studied in several ways. We first present in Fig. \ref{fig:hyperparameters} representative results for the estimation of the MAP model hyper-parameters $(\theta_1, \theta_2, D)$, jointly estimated with the relay function identification under ICM. We present these estimates versus the ICM iterations $j = 1,\ldots,50$. Two important features are evident, the first that the results converge to a set of optimal values and secondly that this occurs relatively rapidly, with very few iterations of ICM required. This is characteristic of all the examples we tested.

\begin{figure}
  \centering
  \subfloat[Approach I: perfect CSI]
{\label{fig:abs_1}
\includegraphics[width=45mm, height=60mm]
{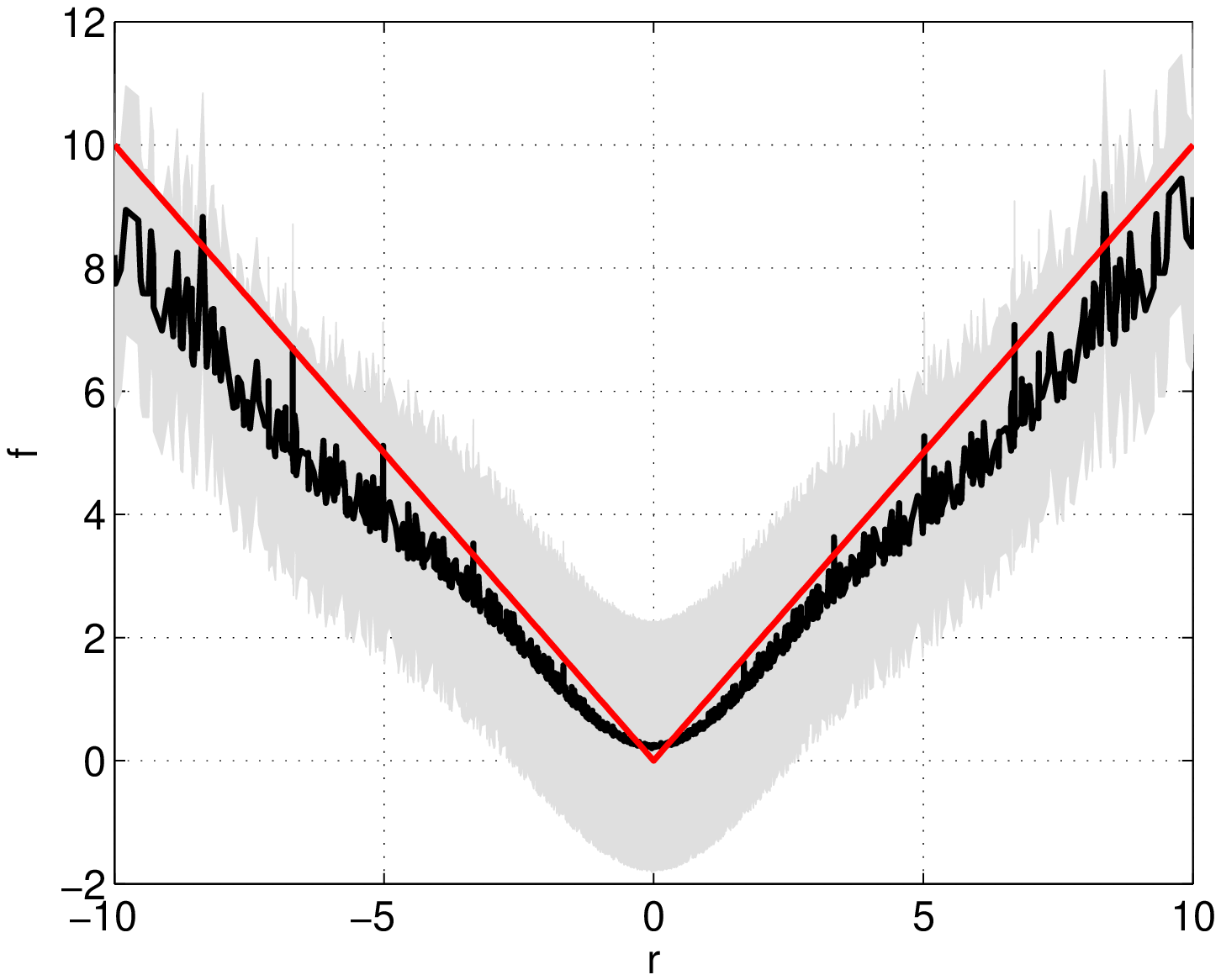}}  
\subfloat[Approach I: imperfect CSI] 
 {\label{fig:abs_2} 
\includegraphics[width=45mm, height=60mm]
{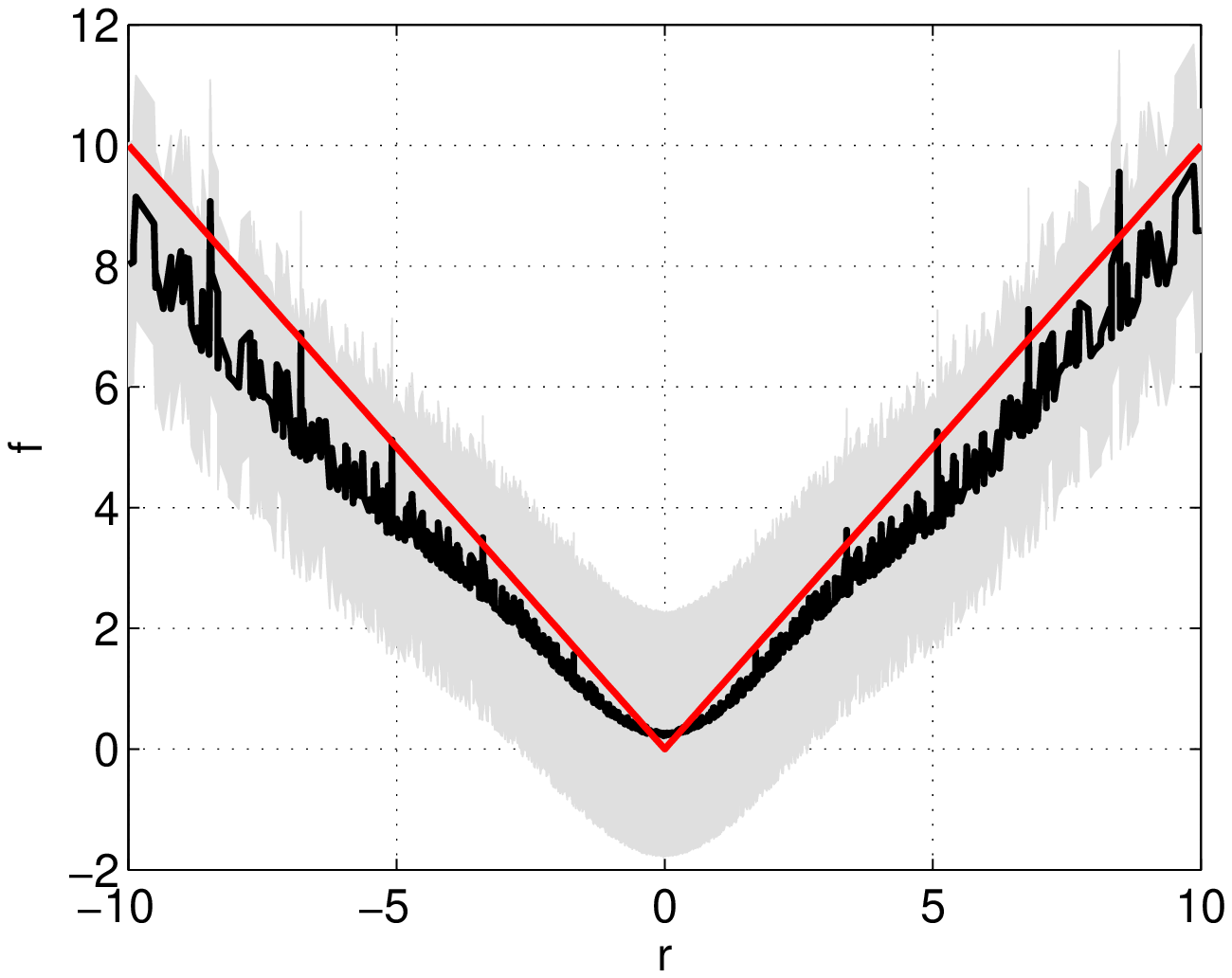}}
\subfloat[Approach II: perfect CSI]
  {\label{fig:abs_3}
  \includegraphics[width=45mm, height=60mm]
  {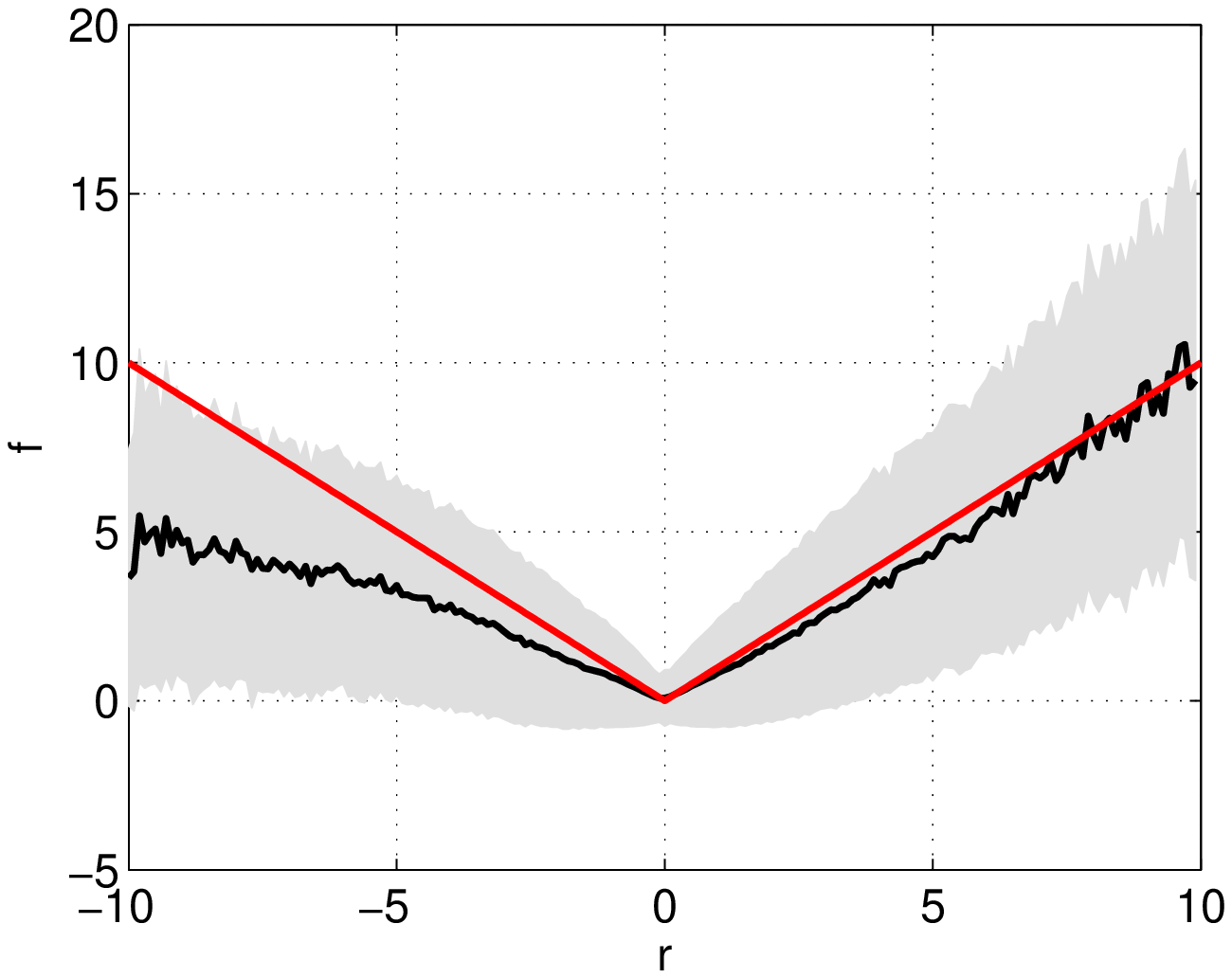}}
\subfloat[Approach II: imperfect CSI]
  {\label{fig:abs_4}
  \includegraphics[width=45mm, height=60mm]
  {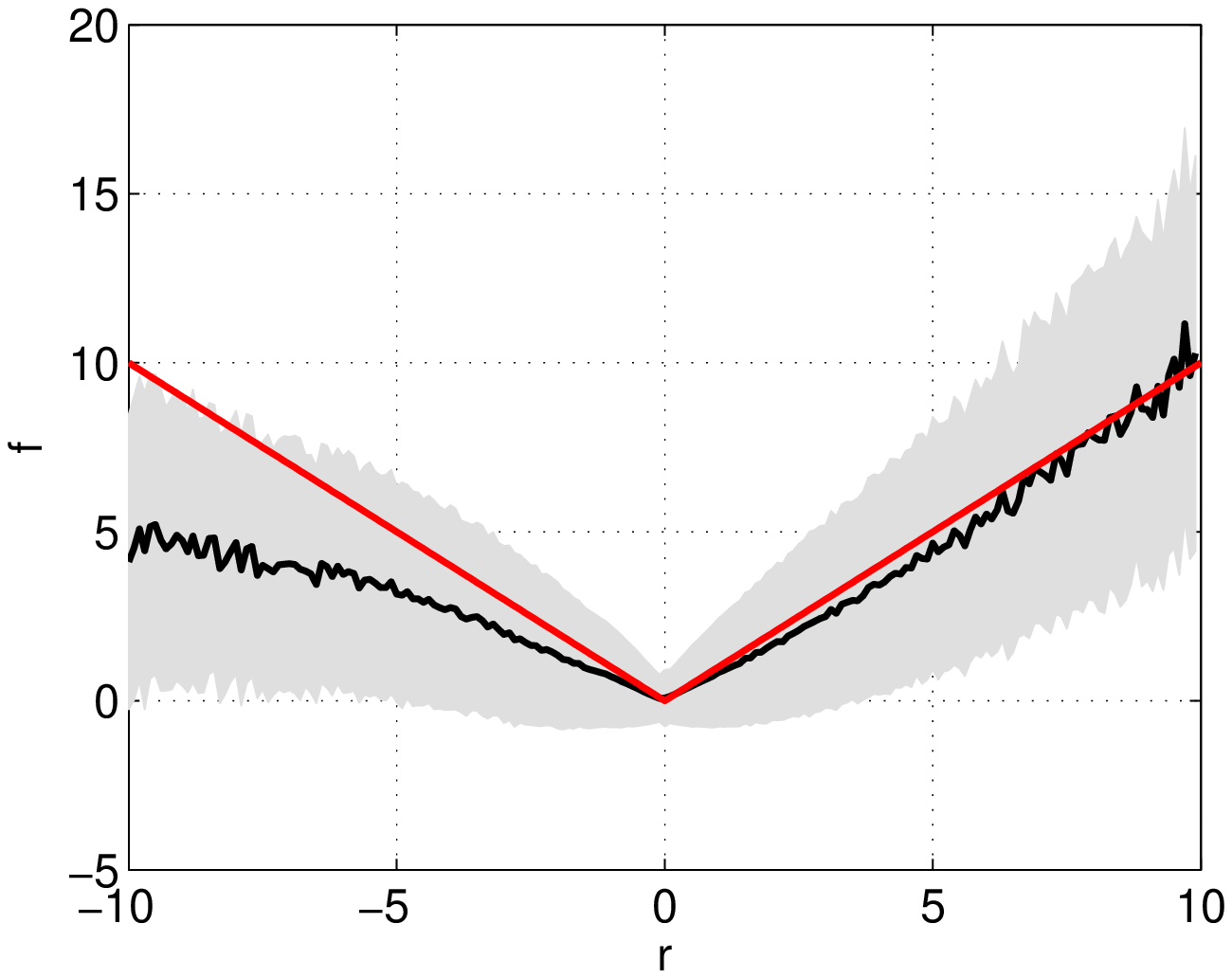}}
    \caption{ABS relay function}
  \label{fig:abs}
\end{figure}

\begin{figure}
  \centering
  \subfloat[Approach I: perfect CSI]
{\label{fig:linear_1}
\includegraphics[width=45mm, height=60mm]
{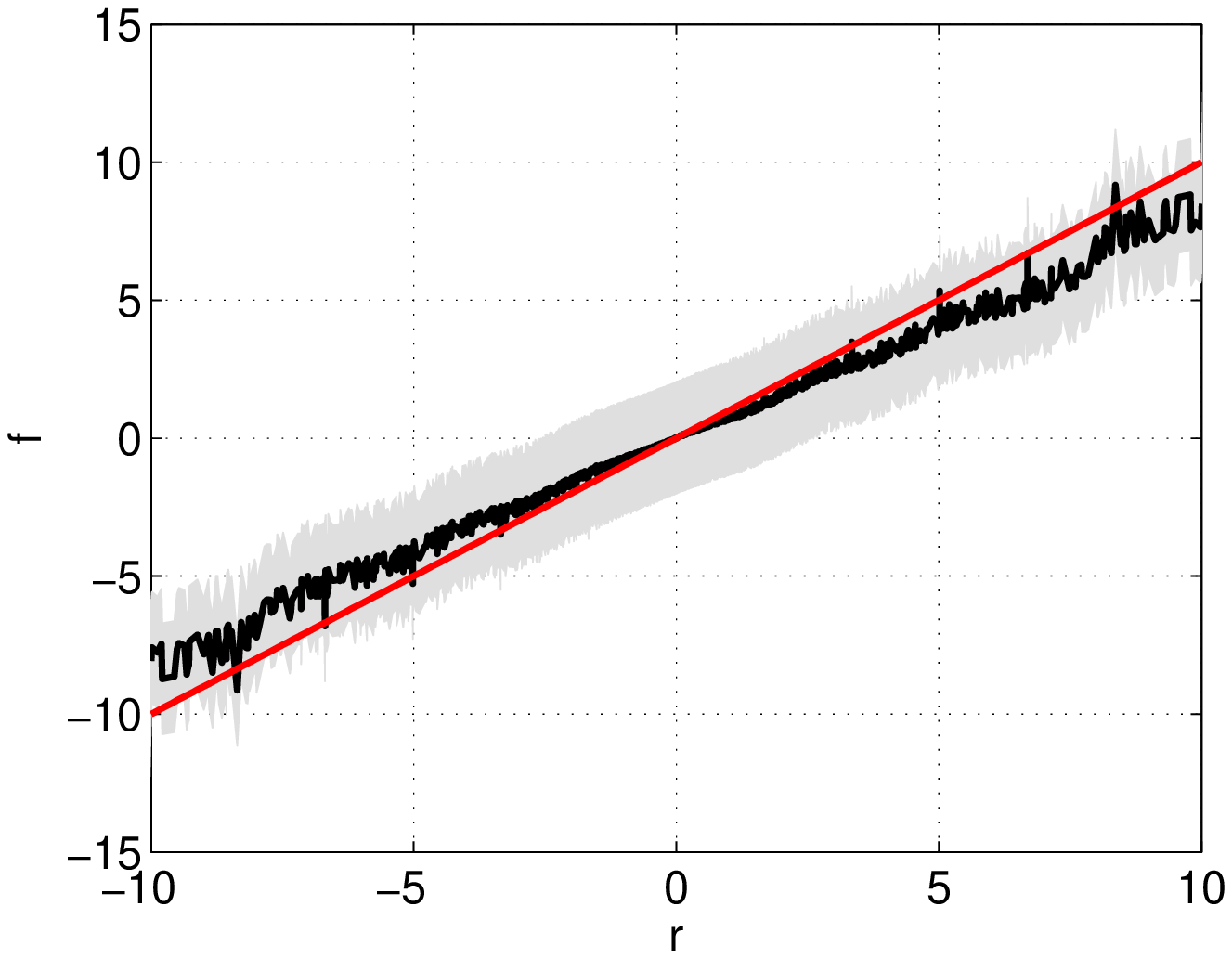}}       
 \subfloat[Approach I: imperfect CSI]
 {\label{fig:linear_2}
 \includegraphics[width=45mm, height=60mm]
 {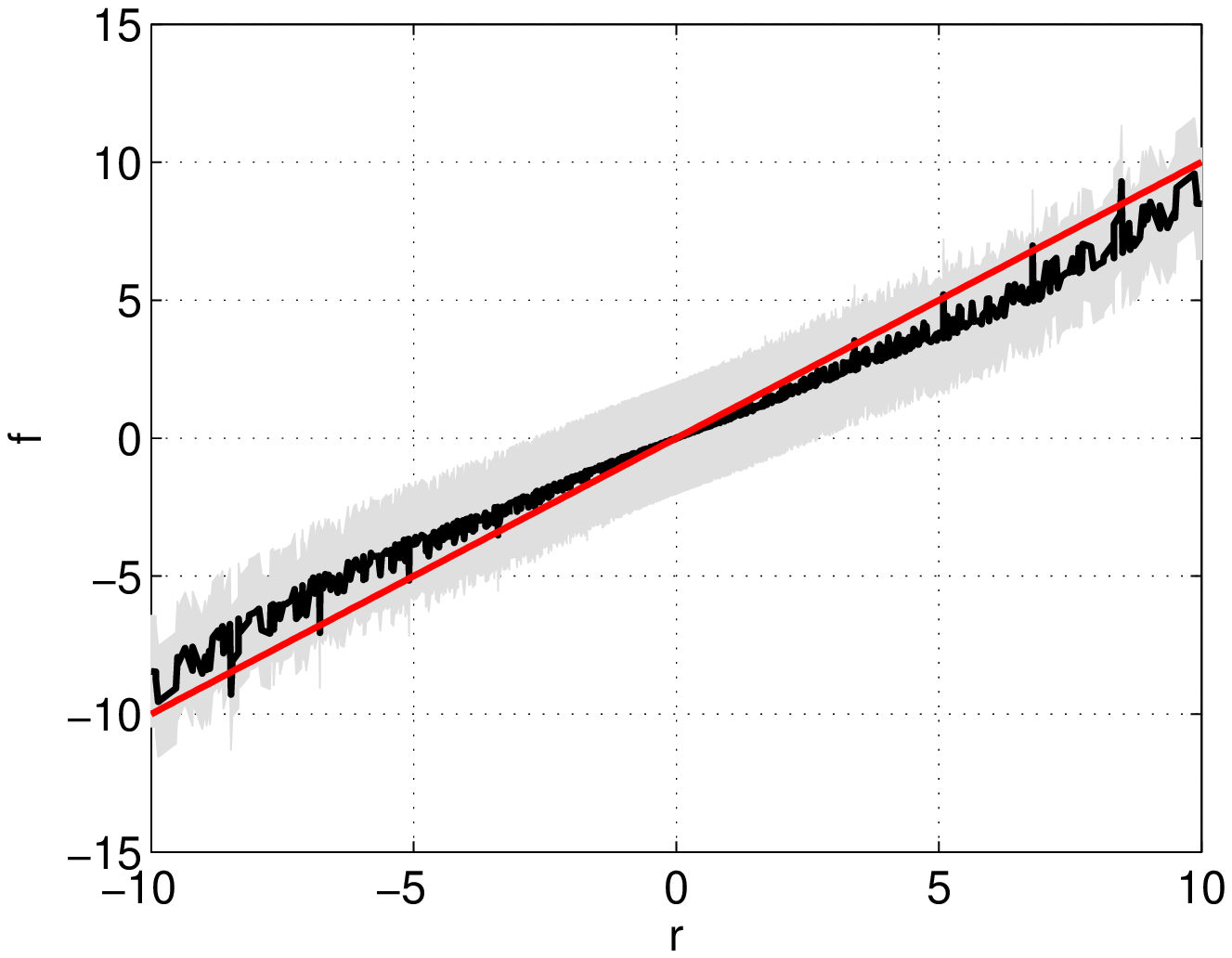}}
\subfloat[Approach II: perfect CSI]
  {\label{fig:linear_3}
  \includegraphics[width=45mm, height=60mm]
  {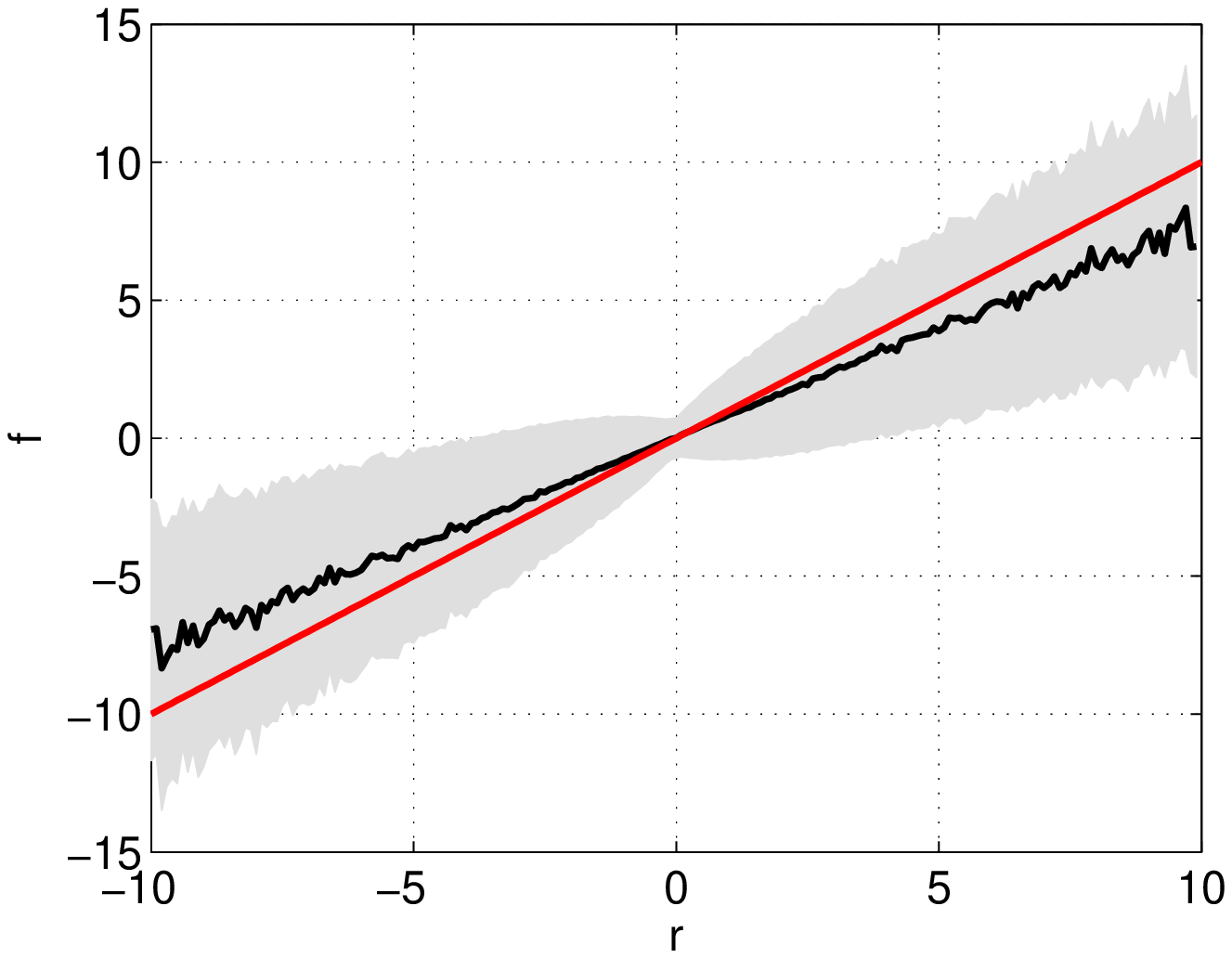}}
\subfloat[Approach II: imperfect CSI]
{\label{fig:linear_4}
\includegraphics[width=45mm, height=60mm]
{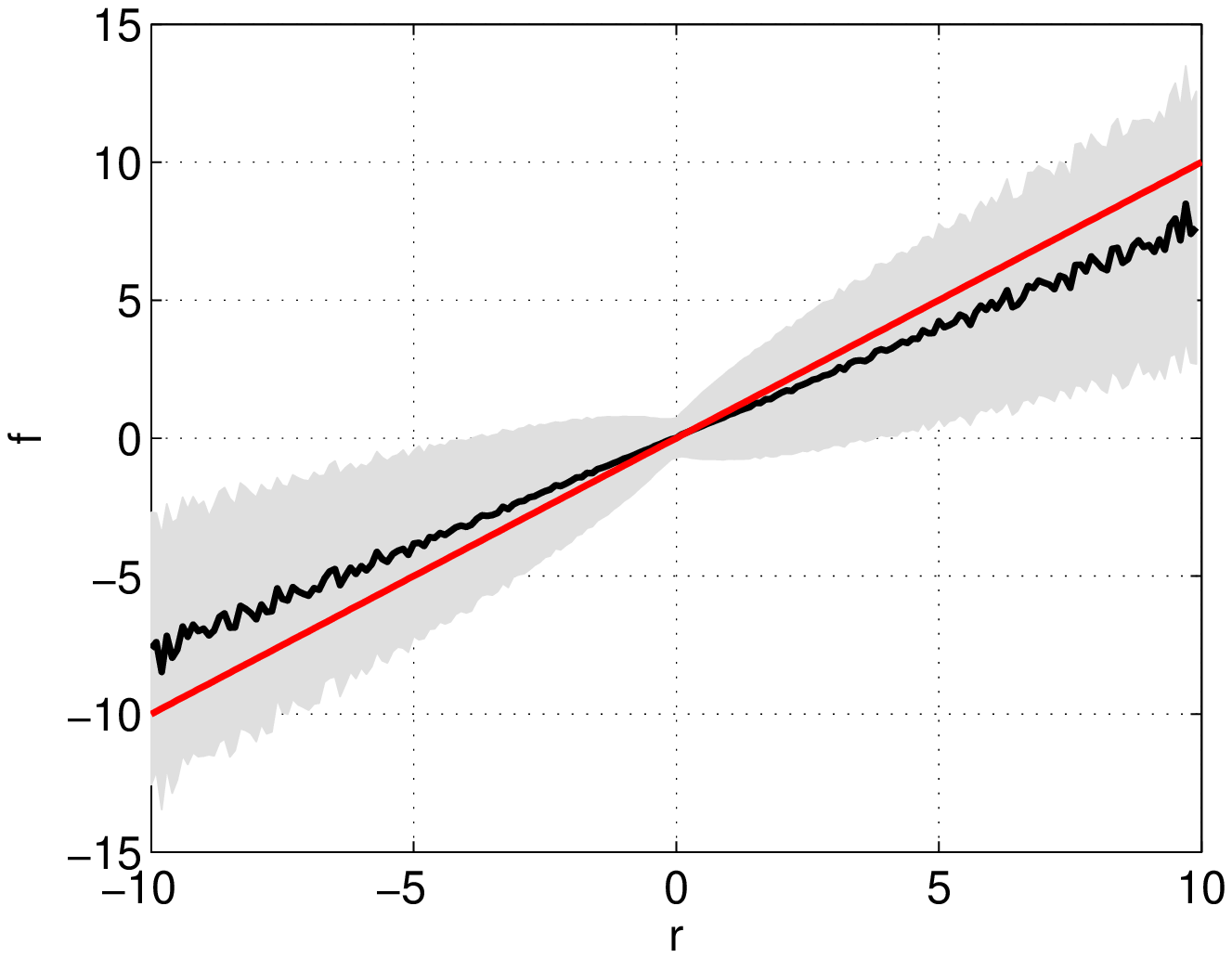}}
    \caption{LINEAR relay function}
  \label{fig:linear}
\end{figure}

\begin{figure}
  \centering
  \subfloat[Approach I: perfect CSI]
  {\label{fig:tanh_1}
  \includegraphics[width=45mm, height=60mm]
  {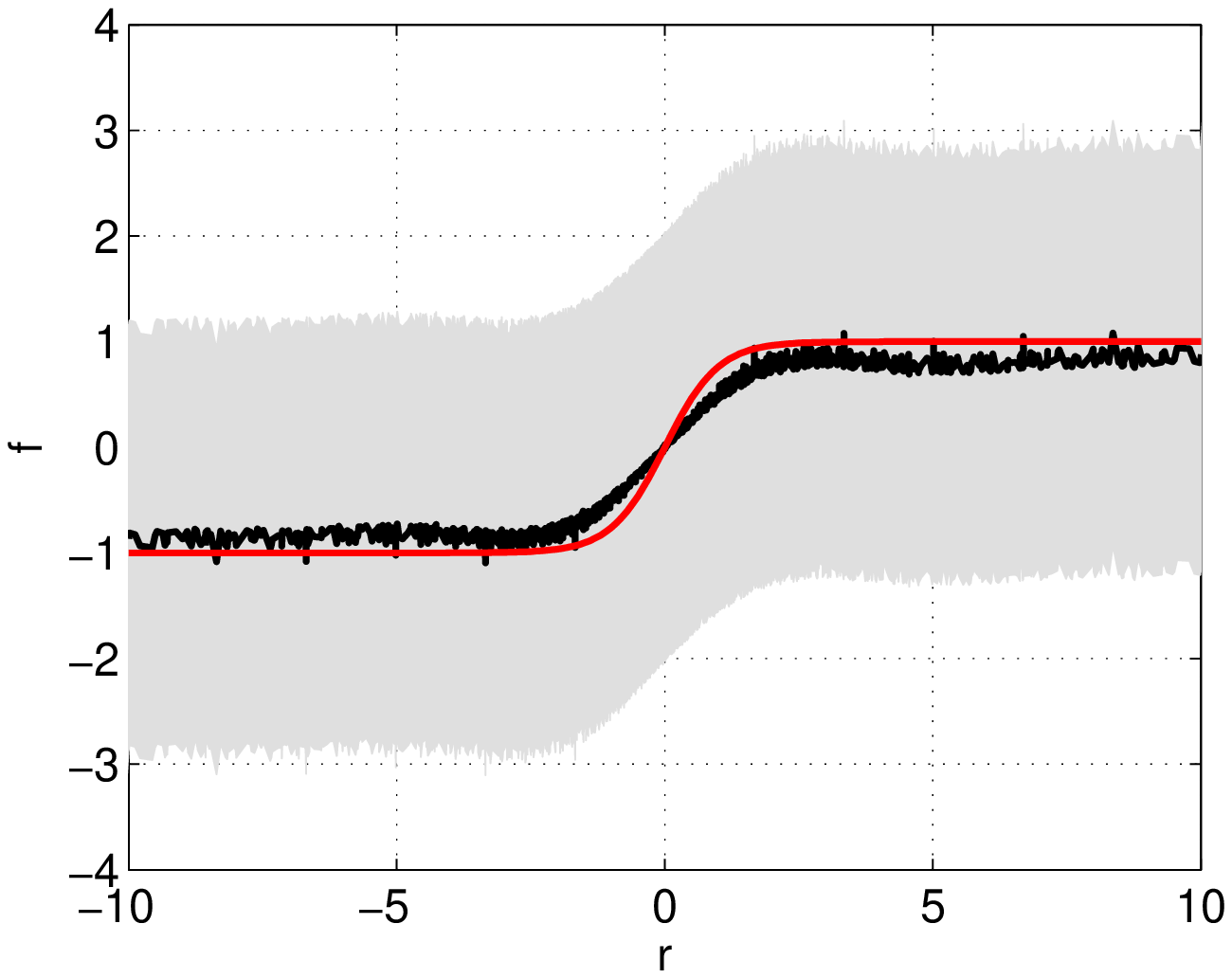}}
  \subfloat[Approach I: imperfect CSI]
  {\label{fig:tanh_2}
  \includegraphics[width=45mm, height=60mm]
  {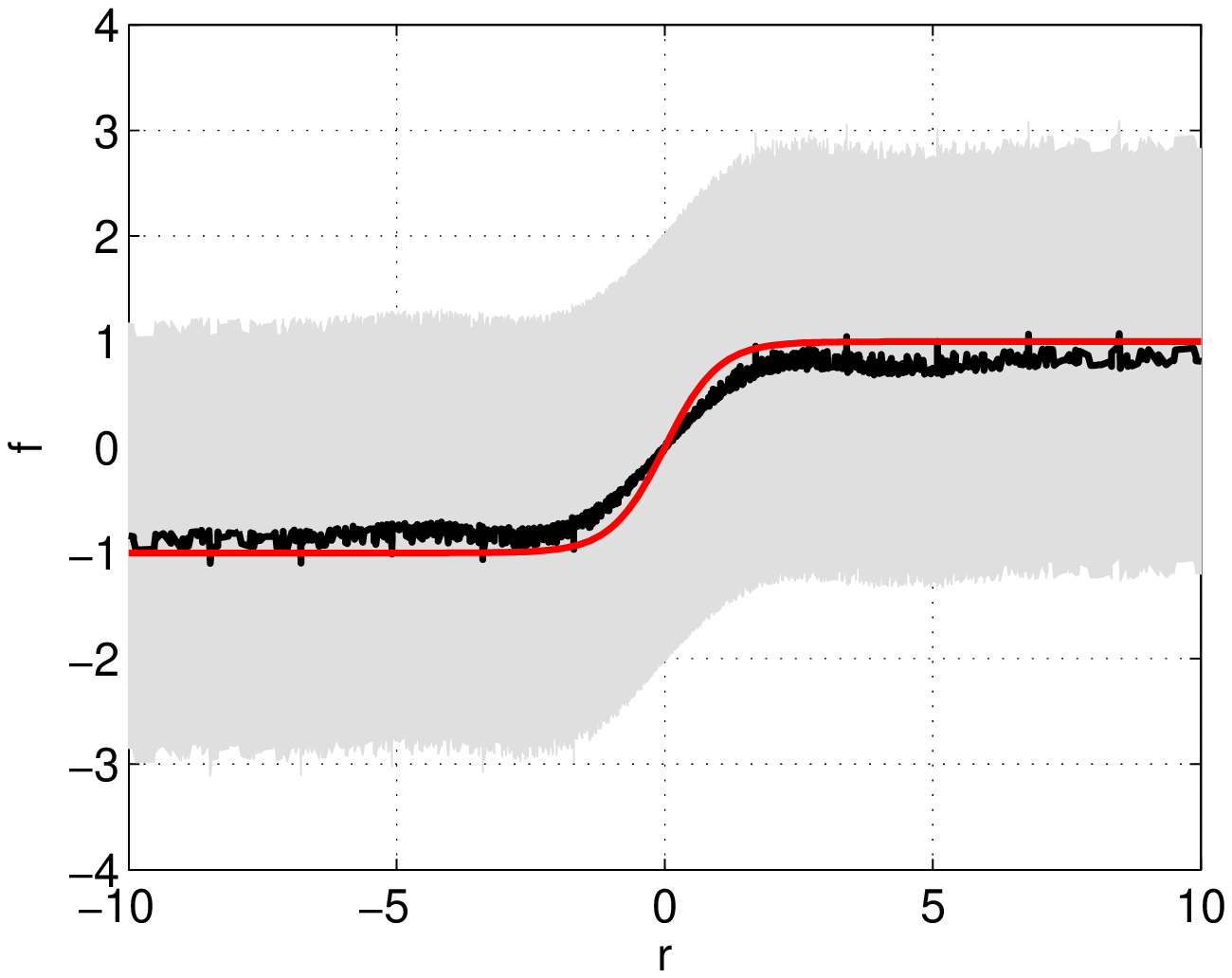}}
  \subfloat[Approach II: perfect CSI]
  {\label{fig:tanh_3}
  \includegraphics[width=45mm, height=60mm]
  {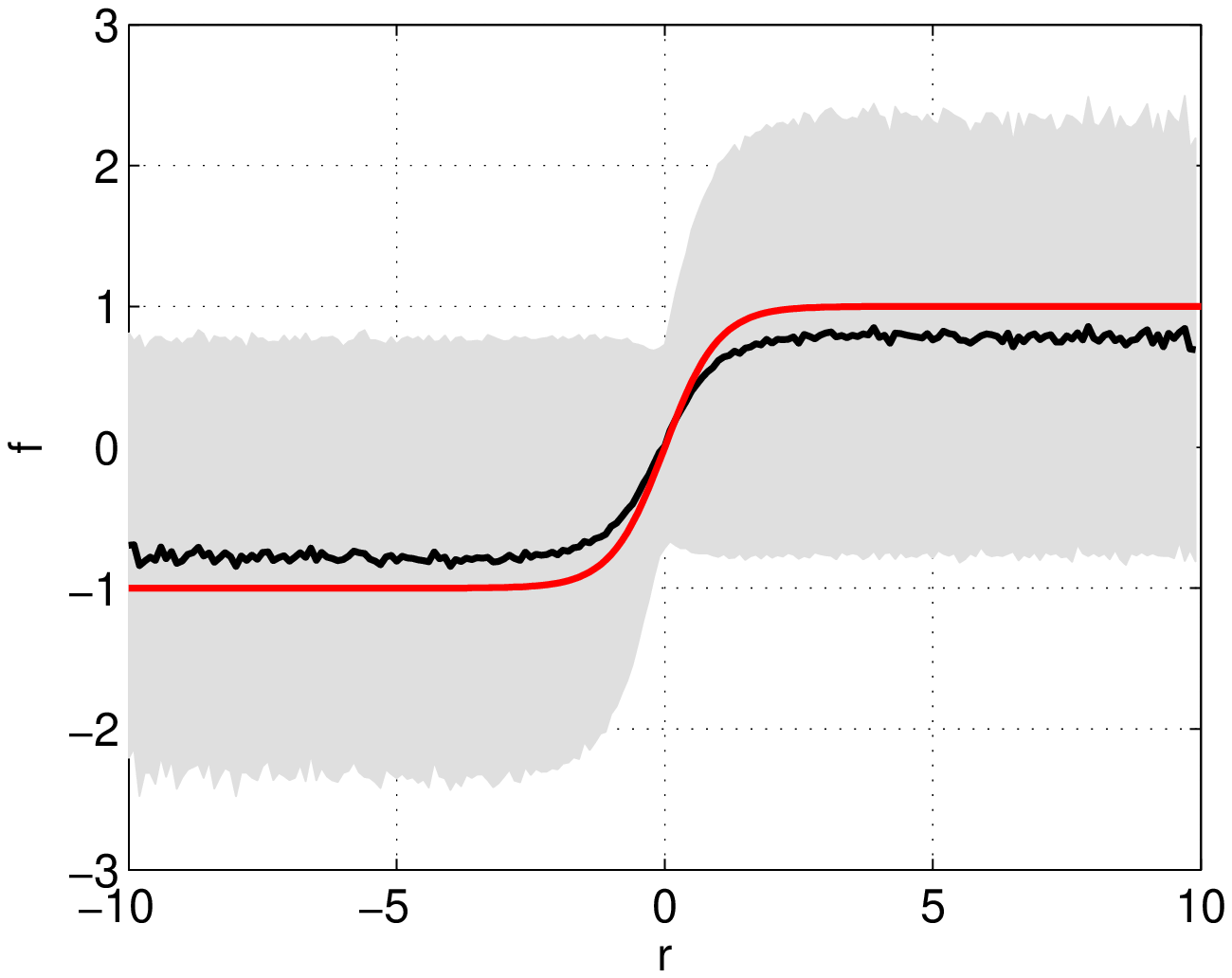}}
  \subfloat[Approach II: imperfect CSI]
  {\label{fig:tanh_4}
  \includegraphics[width=45mm, height=60mm]
  {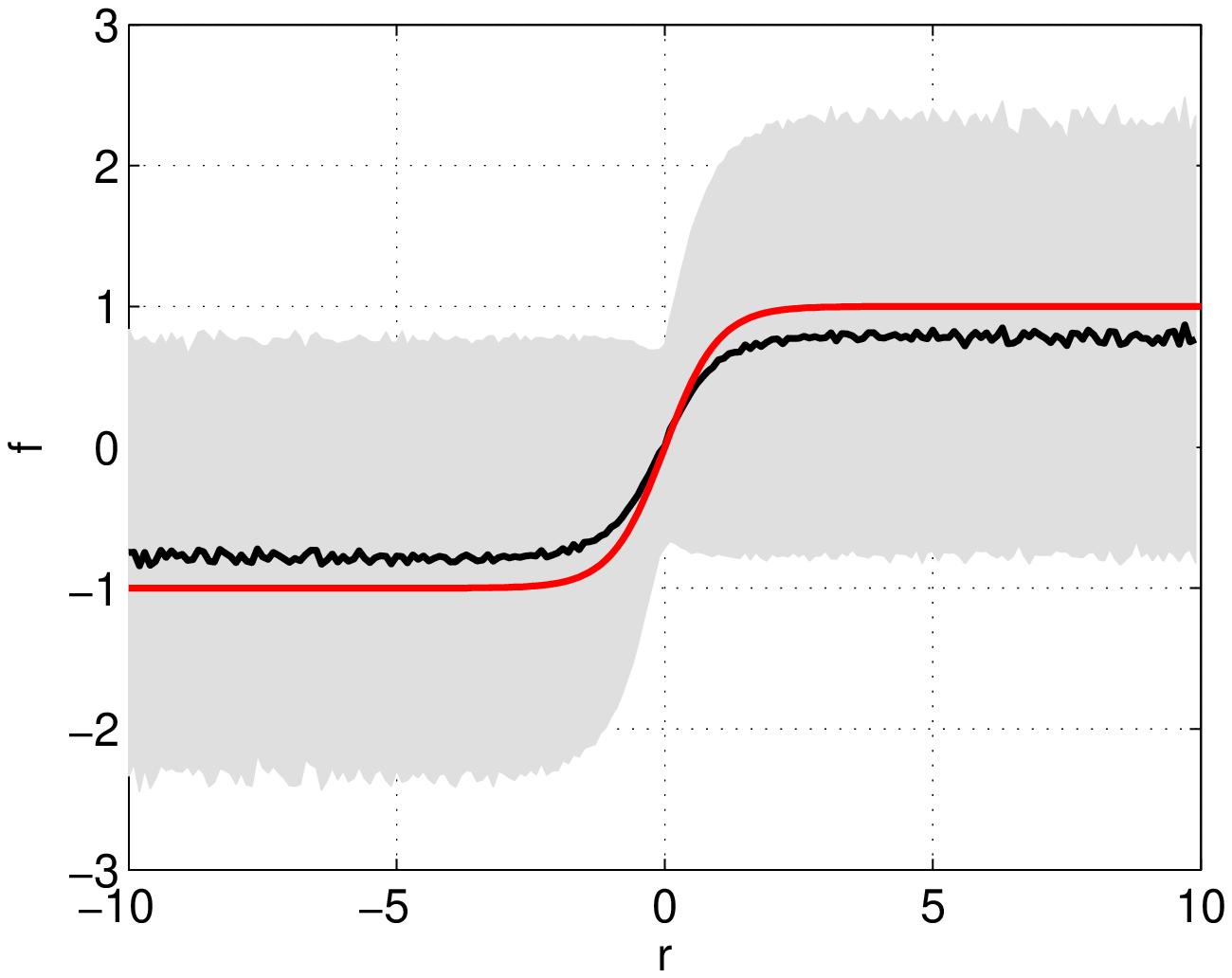}}
  \caption{TANH relay function}
  \label{fig:tanh}
\end{figure}

\begin{figure}
  \centering
  \subfloat[Approach I: perfect CSI]
  {\label{fig:dem_1}
  \includegraphics[width=45mm, height=60mm]
  {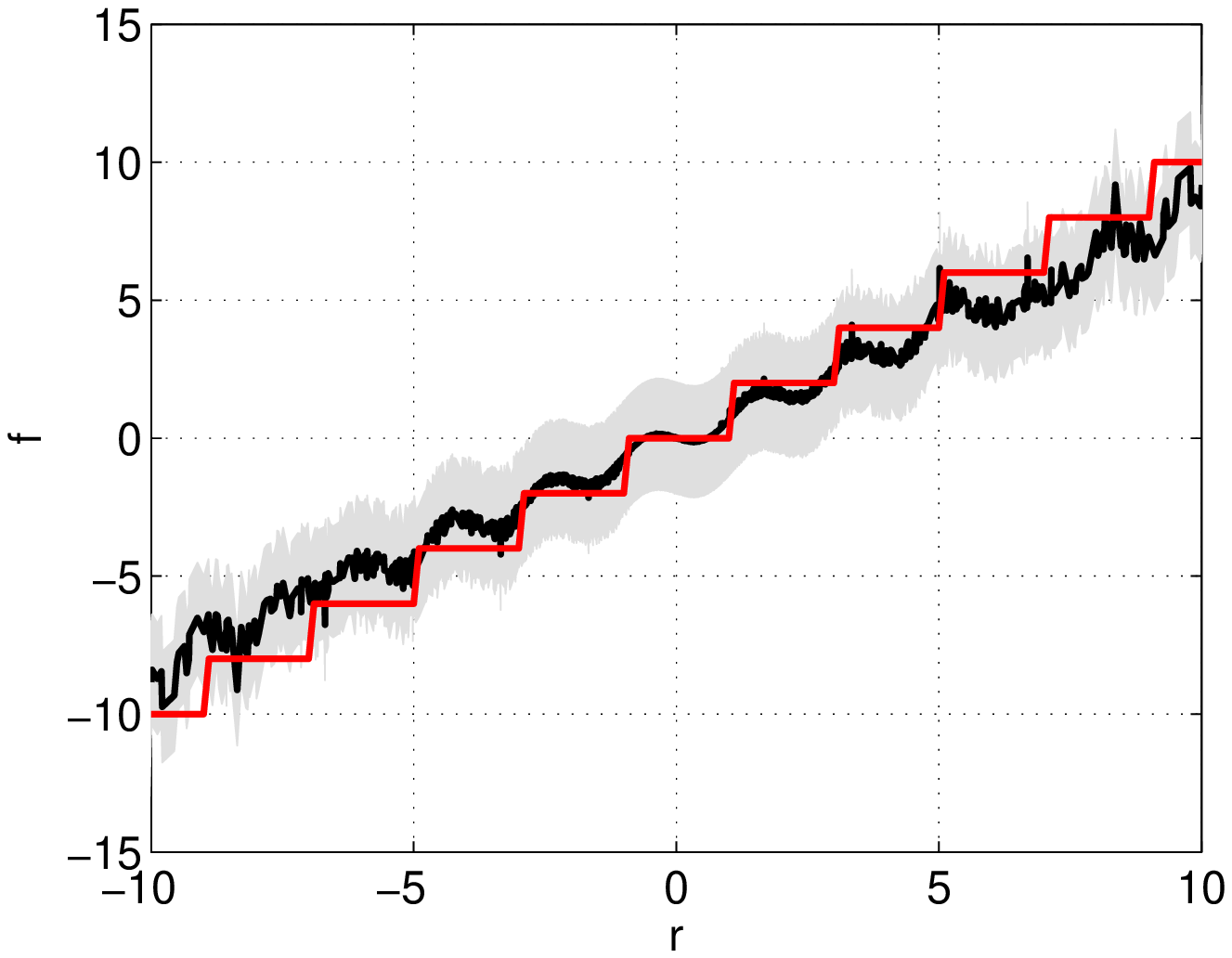}}
  \subfloat[Approach I: imperfect CSI]
  {\label{fig:dem_2}
  \includegraphics[width=45mm,height=60mm]
  {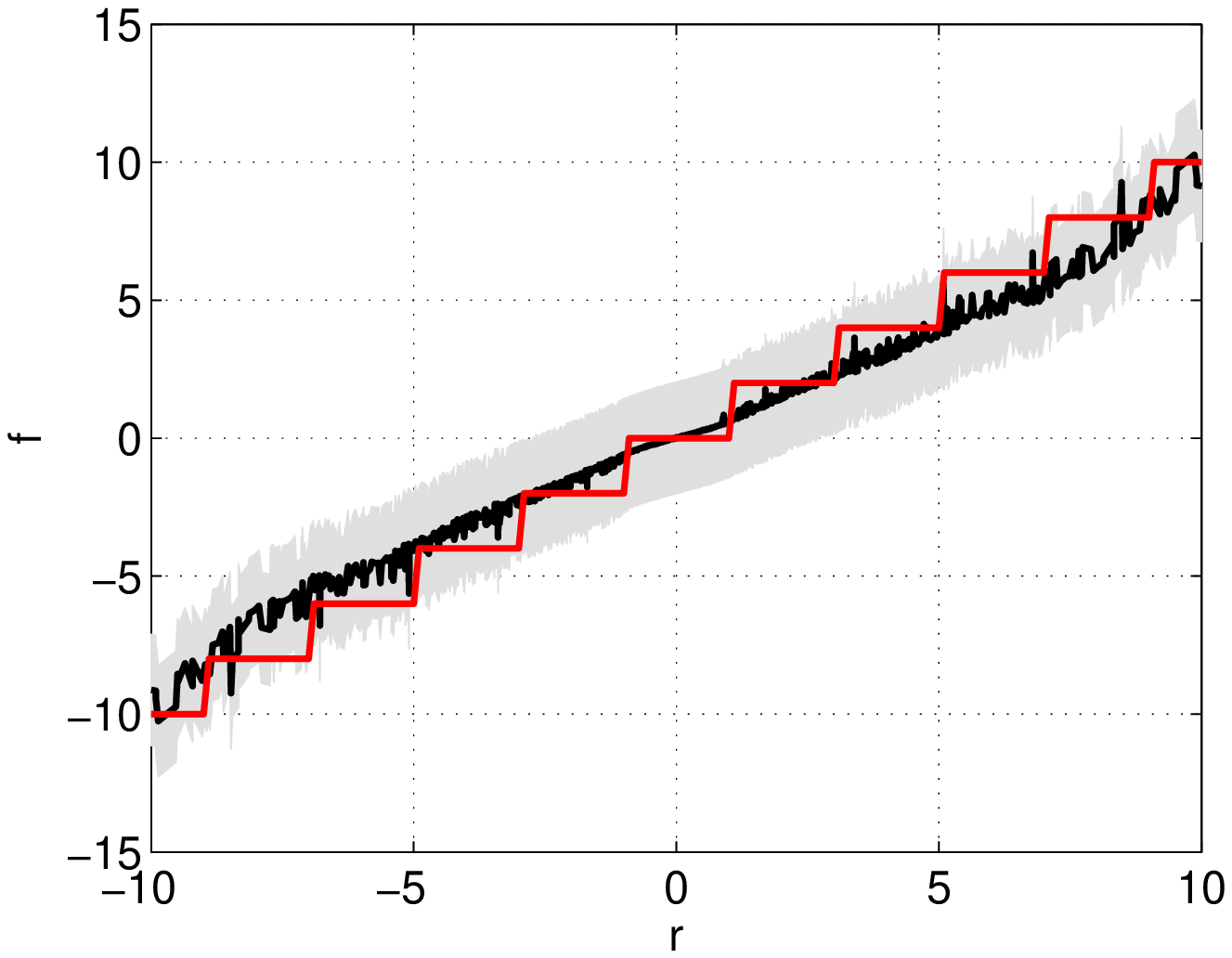}}
  \subfloat[Approach II: perfect CSI]
  {\label{fig:dem_3}
  \includegraphics[width=45mm, height=60mm]
  {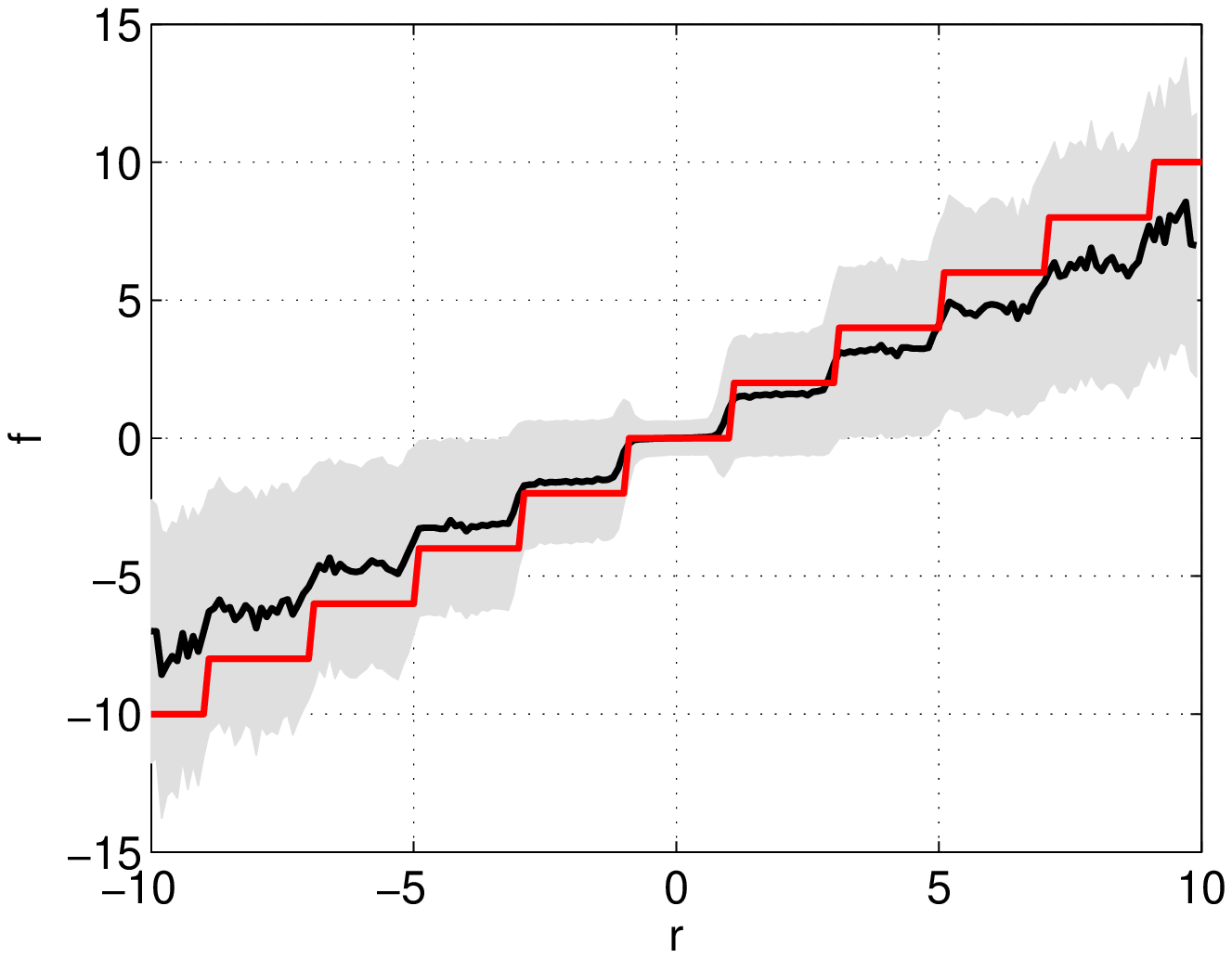}}
  \subfloat[\footnotesize{Approach II: imperfect CSI}]
  {\label{fig:dem_4}
  \includegraphics[width=45mm, height=60mm]
  {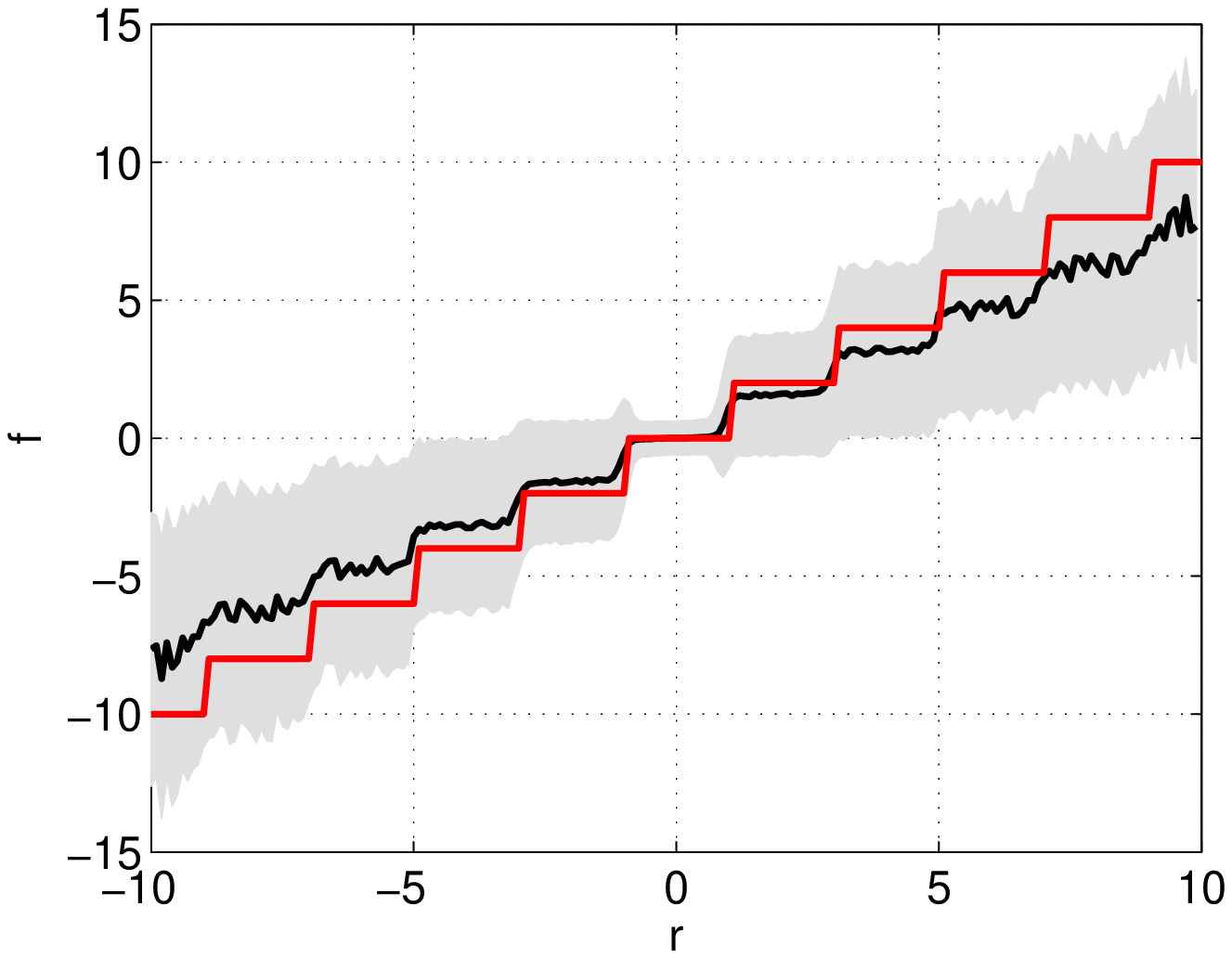}}
  \caption{Demodulate relay function}
  \label{fig:dem}
\end{figure}

Next we summarize the findings for the perfect CSI and imperfect CSI studies of each of the different relay functions: absolute function in Fig. \ref{fig:abs}, linear function in Fig. \ref{fig:linear}, tanh function in Fig. \ref{fig:tanh} and demodulated function in Fig. \ref{fig:dem}. These results are presented each in four sub-plots, the first two are obtained from the Full Information (Approach I) and the last two are for the  Frame-by-Frame (Approach II). The results we present are the posterior MAP via the GP estimation in (\ref{GP_estimation}). In addition, in grey we present the 95\% posterior confidence intervals on these MAP estimates via (\ref{GP_estimation_error}). We compare these estimates to the true relay functional form utilised in the simulations.

\begin{itemize}
\item{For all the non-linear relay functions considered, the estimation under both perfect and imperfect CSI is highly accurate for the Full Information (Approach I).}
\item{The demodulated relay function which had the linear trend with stairs overlayed, was most difficult to perform system identification as it contained a global feature of the linear trend as well as local fine scale features corresponding to the stairs function. We observed that in the Full Information case with perfect CSI, the estimation was relatively accurate. However, for the imperfect CSI settings, the estimation of the global feature of the trend was evident, though the resolution of the local features of the stairs was diminished. Therefore, learning such intricate features will require many more frames of estimation. The results presented were for $100$ frames. With an increase over time to $500$ or more, the estimation will resolve these local features.}
\item{As expected in any regression based anlaysis, the functional forms were most difficult to estimate at the extremities of the convex hull of the received PAM constellation points. This can be shown to result in the largest predictive uncertainty in the estimated function, leading in this case to most uncertainty in the estimated relay functional forms. This was observed in all settings and for all functions, and is most poignant in the Demodulated function example.}
\item{In the Frame-by-Frame results we evaluate the function at fixed grid points set by the constellation transmission symbols space. In this case we observe a smoothing of the estimates which can help resolve the local resolution when compared to the estimation of the relay function at each observed constellation point as undertaken in Approach I. This smoothing approach could also be applied to Approach 1.}
\end{itemize}

Next, we present summarised results in Table \ref{tab:AbsError}, which are each based on the absolute error in the MAP estimated relay functions under Approaches I, II and III over $100$ frames. We observe that, as expected, there is a clear trade-off between computationally efficiency and accuracy in the estimation. The Full-Information (Approach I) is the most accurate. The Overlapping-Sliding-Window, which in this simulation had a $0\%$ overlap, was the least accurate, though it was the most computationally efficient approach corresponding to Frame-by-Frame estimation.

\begin{figure}
    \centering
        \epsfysize=12cm
        \epsfxsize=15cm
        \epsffile{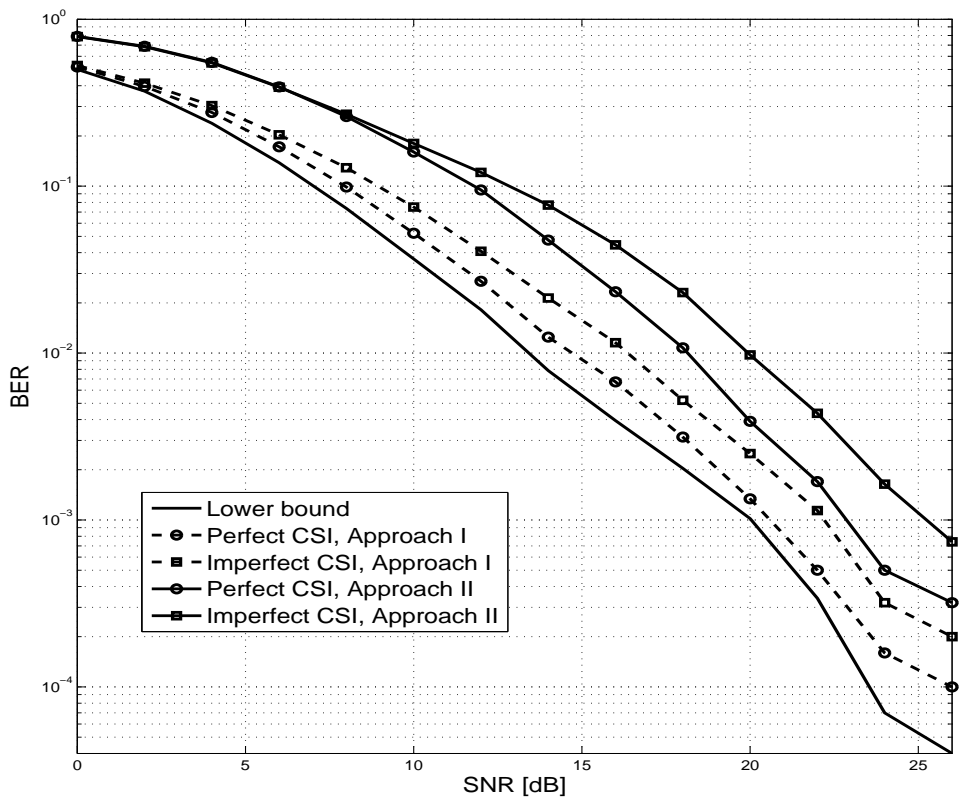}
    \caption{BER for Amplify-and-Forward relay estimation under perfect/imperfect CSI }
  \label{fig:BER}
\end{figure}

Finally we present Bit Error Rate (BER) results for the case of linear relay function. The results are presented in Fig. \ref{fig:BER} for both Full-Information (Approach I) and Frame-by-Frame (Approach II). For comparison, we also plot the BER for the case of perfect CSI and perfect knowledge of the relay function which serves as a lower bound on the BER performance. As the figure shows, there is less than $1$dB difference between the lower bound and Approach I under perfect CSI; and less than $3$ dB difference between the lower bound and Approach II under perfect CSI. In the imperfect CSI case there is a further $2$ dB degradation in both cases. These results demonstrate the importance of accurate estimation of the relay function.
\section{Conclusions} \label{conclusions}
We considered the problem of relay identification in wireless relay systems. We developed a flexible stochastic model for a class of cooperative wireless relay networks, in which the relay processing functionality is not known at the destination. Working under this modelling framework we developed and demonstrated the performance of our estimation procedure aimed at performing efficient system identification. 

In particular we demonstrated that Gaussian Process modelling via ICM approach can resolve the problem of system identification in a computationally efficient algorithm for many different relay functional forms which have desirable characteristics with respect to transmission functionality and quality of service.

\bibliographystyle{IEEEtran}
\bibliography{../../../../references}

\section{Appendix}
Here we provide a proof for the expressions in Theorem \ref{Theorem_1}.

\begin{proof}
The posterior distribution decomposes according to (\ref{PosteriorModel}), and we can therefore derive the following quantities for a given relay
\begin{enumerate}
\item The full conditional posterior distribution for the $l$-th relay function in (\ref{posterior_1}) is given by
\begin{equation}
p\left( \f^{(l)}_{1:T} | \THETA^{(l)},D^{(l)},\y^{(l)}_{1:T} \right)
=\frac 
{p\left(\y^{(l)}_{1:T}|\f^{(l)}_{1:T}\right)
p \left(\f^{(l)}_{1:T};\mu^{(l)}_{\THETA^{(l)}}\left(\r_k^{(l)}(t) \right),\mathcal{K}^{(l)}_{1:T}\right)}
{
\int   p\left(\y^{(l)}_{1:T}|\f^{(l)}_{1:T}\right) 
p \left(
\f^{(l)}_{1:T};\mu^{(l)}_{\THETA^{(l)}}\left(\r_k^{(l)}(t) \right),\mathcal{K}^{(l)}_{1:T}
\right) d \f^{(l)}_{1:T} }.
\end{equation}
With a matrix variate normal likelihood model for $\y^{(l)}_{1:T} | \f^{(l)}_{1:T}$ and a GP prior on the function will result in a matrix variate prior for the function over the symbols in each frame, i.e $K \times T$.
After vectorizing the observation random matrix and the prior random matrix, we obtain multi variate Gaussian distributions which admit standard conjugacy properties, see \cite{fink1997compendium}. 
This results in 
\begin{equation}
p\left( \text{Vec}\left[\f^{(l)}_{1:T} \right]| \THETA^{(l)},D^{(l)},\y^{(l)}_{1:T} \right)
= N\left(\M, \Sigma \right),
\end{equation}
where $\M$ and $\Sigma$ are defined in (\ref{MeanFunction}-\ref{Conjugate_2}).
	\item The full conditional for $\THETA^{(l)}$ in (\ref{posterior_2}) can be expressed as
{\small{
\begin{equation}
\begin{split}
&p\left( \THETA^{(l)}|\D^{(l)}, \f^{(l)}_{1:T}\left(\cdot\right),\y^{(l)}_{1:T} \right)\\
&\propto
p\left(\text{Vec}\left[\y_{1:T}^{(l)}\right]|\text{Vec}\left[\f^{(l)}_{1:T}\right],\THETA^{(l)},D^{(l)}\right)
p \left(\text{Vec}\left[\f^{(l)}_{1:T}\right];\bm{\mu}^{(l)}_{1:T},\mathcal{K}^{(l)}_{1:T}\right)
p\left(\THETA^{(l)}\right)\\
&=
N \left(\text{Vec}\left[\f^{(l)}_{1:T}\right];\bm{\mu}^{(l)}_{1:T},\mathcal{K}^{(l)}_{1:T}\right)
p\left(\THETA^{(l)}\right).
\end{split}
\end{equation}}}
Deriving the MAP estimate is achieved by applying the $\log$ transform to the posterior, taking partial derivative with respect to each parameter, setting to $0$ and solving as follows:
{\small{\begin{equation}
\begin{split}
\large 
{\triangledown _{\THETA^{(l)}}}
\left[-\log \left(2 \pi \det \left[\mathcal{K}^{(l)}_{1:T}\right]\right)^{\frac{K\times T}{2}}
- \frac{1}{2}  
\left(\text{Vec}\left[\f^{(l)}_{1:T}\right]-\bm{\mu}^{(l)}_{1:T} \right)^T
\left(\mathcal{K}^{(l)}_{1:T}\right)^{-1}
\left(\text{Vec}\left[\f^{(l)}_{1:T}\right]-\bm{\mu}^{(l)}_{1:T} \right)
+ \log p\left(\THETA^{(l)}\right)\right] = \bm{0}
\end{split}
\end{equation}}}
gives
{\small{
\begin{equation}
\begin{split}
\large 
{\triangledown _{\THETA^{(l)}} }
&
\left[\left(\text{Vec}\left[\f^{(l)}_{1:T}\right]-\bm{\mu}^{(l)}_{1:T} \right)^T
\left(\mathcal{K}^{(l)}_{1:T}\right)^{-1}
\left(\text{Vec}\left[\f^{(l)}_{1:T}\right]-\bm{\mu}^{(l)}_{1:T} \right)
+ \log p\left(\THETA^{(l)}\right)\right]  = \\
&
\left[
\begin{array}{cccccccccc}
 1 &  1 &1 &1 &1 &1 &1 &1 &1 &1    \\
 r_1^{(l)}(1) &  \cdots    & r_K^{(l)}(1) & 
 r_1^{(l)}(2) &  \cdots    & r_K^{(l)}(2) & 
 \cdots&
 r_1^{(l)}(T) &  \cdots    & r_K^{(l)}(T)  
\end{array}
\right]\\
&\times
\left(\text{Vec}\left[\f^{(l)}_{1:T}\right]^T\left(\mathcal{K}^{(l)}_{1:T}\right)^{-1}
+2 \left(\bm{\mu}^{(l)}_{1:T}\right)^T\left(\mathcal{K}^{(l)}_{1:T}\right)^{-1}\right)^T
+2 \left(\bm{\Theta}^{(l)}\right)^T\left(\Sigma_{\bm{\Theta}}\right)^{-1} =\bm{0}.
\end{split}
\end{equation}
}}
It produces the following linear system of equations with a unique solution:
\small
\begin{equation}
\begin{split}
&\left(\theta_1+\theta_2\right)
\sum_{i=1}^K
\sum_{t=1}^T
r_i(t) \left[\left(\mathcal{K}_{1:T}\right)^{-1}\right]_{i ,t}+
\theta_1 \left[\left(\Sigma_{\bm{\Theta}}\right)^{-1}\right]_{1, 1}+
\theta_2 \left[\left(\Sigma_{\bm{\Theta}}\right)^{-1}\right]_{2 ,1}
= 
-\frac{1}{2}
\sum_{i=1}^K
\sum_{t=1}^T
f(r_i(t))
\left[\left(\mathcal{K}_{1:T}\right)^{-1}\right]_{i ,t},\\
&\left(\theta_1+\theta_2\right)
\sum_{i=1}^K
\sum_{t=1}^T
r^2_i(t) \left[\left(\mathcal{K}_{1:T}\right)^{-1}\right]_{i ,t}+
\theta_1 \left[\left(\Sigma_{\bm{\Theta}}\right)^{-1}\right]_{1, 2}+
\theta_2 \left[\left(\Sigma_{\bm{\Theta}}\right)^{-1}\right]_{2 ,2}
= 
-\frac{1}{2}
\sum_{i=1}^K
\sum_{t=1}^T
r_i(t) f(r_i(t))
\left[\left(\mathcal{K}_{1:T}\right)^{-1}\right]_{i ,t},
\end{split}
\end{equation}
\normalsize
giving

\begin{equation}
\begin{split}
&\theta_1 = \frac
{-\theta_2\left(
\left[\left(\Sigma_{\bm{\Theta}}\right)^{-1}\right]_{2 ,1}+
\sum_{i=1}^K\sum_{t=1}^Tr_i(t) \left[\left(\mathcal{K}_{1:T}\right)^{-1}\right]_{i ,t}  \right)
-\frac{1}{2} \sum_{i=1}^K \sum_{t=1}^T f(r_i(t)) \left[\left(\mathcal{K}_{1:T}\right)^{-1}\right]_{i ,t}}
{
\left[\left(\Sigma_{\bm{\Theta}}\right)^{-1}\right]_{1 ,1}+
\sum_{i=1}^K\sum_{t=1}^T r_i(t) \left[\left(\mathcal{K}_{1:T}\right)^{-1}\right]_{i ,t}  
},\\
&\theta_2 = \frac{1}{2 \Phi}
\left(
-\sum_{i=1}^K\sum_{t=1}^T r_i(t)  f(r_i(t)) \left[\left(\mathcal{K}_{1:T}\right)^{-1}\right]_{i ,t}
\right.\\
&\left.+
\frac
{
\left(
\sum_{i=1}^K\sum_{t=1}^T f(r_i(t)) \left[\left(\mathcal{K}_{1:T}\right)^{-1}\right]_{i ,t}
\right)
\left(
\left[\left(\Sigma_{\bm{\Theta}}\right)^{-1}\right]_{1 ,2}
+
\sum_{i=1}^K\sum_{t=1}^T r^2_i(t) \left[\left(\mathcal{K}_{1:T}\right)^{-1}\right]_{i ,t}
\right)
}
{
\left[\left(\Sigma_{\bm{\Theta}}\right)^{-1}\right]_{1 ,1}+
\sum_{i=1}^K\sum_{t=1}^T r_i(t) \left[\left(\mathcal{K}_{1:T}\right)^{-1}\right]_{i ,t}  
}
\right).
\end{split}
\end{equation}
Here, 
\begin{equation}
\begin{split}
\Phi = 
&\frac{-\left(
\left[\left(\Sigma_{\bm{\Theta}}\right)^{-1}\right]_{2 ,1}+
\sum_{i=1}^K\sum_{t=1}^T r_i(t) \left[\left(\mathcal{K}_{1:T}\right)^{-1}\right]_{i ,t}  
\right)
\left(
\left[\left(\Sigma_{\bm{\Theta}}\right)^{-1}\right]_{1 ,2}+
\sum_{i=1}^K\sum_{t=1}^T r^2_i(t) \left[\left(\mathcal{K}_{1:T}\right)^{-1}\right]_{i ,t}  
\right)
}
{
\left[\left(\Sigma_{\bm{\Theta}}\right)^{-1}\right]_{1 ,1}+
\sum_{i=1}^K\sum_{t=1}^T r_i(t) \left[\left(\mathcal{K}_{1:T}\right)^{-1}\right]_{i ,t}  
}\\
&+
\left[\left(\Sigma_{\bm{\Theta}}\right)^{-1}\right]_{2 ,2}+
\sum_{i=1}^K\sum_{t=1}^T r^2_i(t) \left[\left(\mathcal{K}_{1:T}\right)^{-1}\right]_{i ,t}  .
\end{split}
\end{equation}
\item
The full conditional for $\D^{(l)}$ in (\ref{posterior_3}) can be expressed as
\begin{equation}
\begin{split}
p\left(\D^{(l)} | \f^{(l)}_{1:T},\THETA^{(l)},\y^{(l)}_{1:T} \right)
\propto
N \left(\text{Vec}\left[\f^{(l)}_{1:T}\right];\bm{\mu}^{(l)}_{1:T},\mathcal{K}^{(l)}_{1:T}\right)
p\left(\D^{(l)}\right)
\end{split}
\end{equation}
where we clarify that $\mathcal{K}^{(l)}_{1:T}$ is implicitly a function of $\D^{(l)} = D^{(l)}$, and the kernel choice given in Section \ref{System_model_and_assumptions} item 8 ensures that this parameter is a scalar random variable.

Deriving the MAP estimate is achieved by applying the $\log$ transform to the posterior, taking
derivative with respect to $D^{(l)}$, setting to 0 and solving as follows:
\small
\begin{equation}
\begin{split}
&\frac{d }
{d D}
\left[
-\frac{1}{2} \log \left|\mathcal{K}_{1:T}\right|
-\frac{1}{2}
\left(\text{Vec}\left[\f_{1:T}\right]^T-\bm{\mu}_{1:T}\right)^T
\left(\mathcal{K}_{1:T}\right)^{-1}
\left(\text{Vec}\left[\f_{1:T}\right]^T-\bm{\mu}_{1:T}\right)
+
\log p\left(D^{(l)}\right)
\right]
\mathbb{I}\left[D^{(l)} \in \left(a,b\right)\right]\\
&=
\left[
-\frac{1}{2} \text{Tr}\left(
\mathcal{K}_{1:T}^{-1}\frac{d\mathcal{K}_{1:T}}{d D^{(l)}}
\right)
+ \frac{1}{2}
\left(\text{Vec}\left[\f_{1:T}\right]^T-\bm{\mu}_{1:T}\right)^T
\left(\mathcal{K}_{1:T}\right)^{-1}
\frac{d\mathcal{K}_{1:T}}{d D^{(l)}}
\left(\mathcal{K}_{1:T}\right)^{-1}
\left(\text{Vec}\left[\f_{1:T}\right]^T-\bm{\mu}_{1:T}\right)
\right]
\mathbb{I}\left[D^{(l)} \in \left(a,b\right)\right].
\end{split}
\end{equation}
\normalsize
This expression is obtained by using standard matrix derivative identities.
\end{enumerate}
\end{proof}
\pagebreak

\begin{table}[b]
\begin{tabular}{|c||c|c||c|c||c|c||c|c||c|c||c|c|}
\hline
\multicolumn{1}{|c||}{$\mbox{}$} & 
\multicolumn{4}{c||}{Full Information} &
\multicolumn{4}{c||}{Sliding Window} &
\multicolumn{4}{c|}{Frame-by-Frame}\\
\hline
\multicolumn{1}{|c||}{$\mbox{}$} & 
\multicolumn{2}{c||}{perfect CSI} &
\multicolumn{2}{c||}{imperfect CSI} &
\multicolumn{2}{c||}{perfect CSI} &
\multicolumn{2}{c||}{imperfect CSI} &
\multicolumn{2}{c||}{perfect CSI} &
\multicolumn{2}{c|}{imperfect CSI} 
\\
\hline
Relay & high  & low  & high  & low & high  & low  & high  & low & high  & low  & high  & low  \\
function & SNR & SNR & SNR & SNR& SNR & SNR & SNR & SNR&  SNR & SNR & SNR & SNR \\
\thickhline
ABS & 0.91 & 3.15 & 0.97 & 3.37 &1.26 & 3.65 &1.43  &3.83 &1.29 & 3.69 &1.52  &3.93\\
\hline
LINEAR  & 0.57  & 0.76 &0.76&0.83& 1.07  & 1.24 &1.11&1.31 &1.15  & 1.24 &1.26&1.43\\
\hline
TANH  & 0.18 & 0.21 &0.19&0.22& 0.2 & 0.23 &0.22&0.24&0.28 & 0.31 &0.33&0.42\\
\hline
DEM  & 0.83 & 0.88 &0.85&0.91& 1.11 & 1.32 &1.21&1.42& 1.23 & 1.45 &1.51&1.74\\
\hline
\end{tabular}
\caption{Comparison of absolute error in MAP estimation for relay functions under each estimation approach, for perfect and imperfect CSI.}
\label{tab:AbsError}
\end{table}

\end{document}